\documentclass[letterpaper]{article} 

\ifdefined\aaaianonymous
    \usepackage[submission]{aaai2026}  
\else
    \usepackage{aaai2026}              
\fi

\usepackage{times}  
\usepackage{helvet}  
\usepackage{courier}  
\usepackage[hyphens]{url}  
\usepackage{graphicx} 
\usepackage{natbib}  
\usepackage{caption} 
\usepackage{multirow}
\usepackage{amsmath} 
\usepackage{algorithm}
\usepackage{algorithmic}

\usepackage[table]{xcolor}
\usepackage{multirow}
\usepackage{makecell}

\usepackage{hyperref} 
\usepackage{newfloat}
\usepackage{listings}
\DeclareCaptionStyle{ruled}{labelfont=normalfont,labelsep=colon,strut=off} 
\floatstyle{ruled}
\newfloat{listing}{tb}{lst}{}
\floatname{listing}{Listing}

\nocopyright 

\ifdefined\aaaianonymous
    \title{PET2Rep: Towards Vision-Language Model-Drived Automated Radiology Report Generation for Positron Emission Tomography}
\else
    \title{PET2Rep: Towards Vision-Language Model-Drived Automated Radiology Report Generation for Positron Emission Tomography}
\fi

\author{
Yichi Zhang\textsuperscript{ \rm 1,2,\equalcontrib}, Wenbo Zhang\textsuperscript{\rm 1,2,\equalcontrib}, Zehui Ling\textsuperscript{\rm 1,2,\equalcontrib}
Gang Feng\textsuperscript{\rm 3}, Sisi Peng\textsuperscript{\rm 3}, \\
Deshu Chen\textsuperscript{\rm 1,2}, Yuchen Liu\textsuperscript{\rm 1,2}, Hongwei Zhang\textsuperscript{\rm 1,2}, Shuqi Wang \textsuperscript{\rm 1}, Lanlan Li \textsuperscript{\rm 1}, \\ Limei Han\textsuperscript{\rm 1,2}, Yuan Cheng\textsuperscript{\rm 1,2,\dag}, Zixin Hu\textsuperscript{\rm 1,2,\dag}, Yuan Qi\textsuperscript{\rm 1,2,\dag}, Le Xue\textsuperscript{\rm 1,2,\dag}
}
\affiliations{
    \textsuperscript{\rm 1} Fudan University \textsuperscript{\rm 2} Shanghai Academy of Artificial Intelligence for Science \\
    \textsuperscript{\rm 3} Shanghai Universal Medical Imaging Diagnostic Center \\
}

\usepackage{bibentry}

\begin{document}

\maketitle

\begin{abstract}
Positron emission tomography (PET) is a cornerstone of modern oncologic and neurologic imaging, distinguished by its unique ability to illuminate dynamic metabolic processes that transcend the anatomical focus of traditional imaging technologies. Radiology reports are essential for clinical decision making, yet their manual creation is labor-intensive and time-consuming. Recent advancements of vision-language models (VLMs) have shown strong potential in medical applications, presenting a promising avenue for automating report generation. However, existing applications of VLMs in the medical domain have predominantly focused on structural imaging modalities, while the unique characteristics of molecular PET imaging have largely been overlooked. To bridge the gap, we introduce PET2Rep, a large-scale comprehensive benchmark for evaluation of general and medical VLMs for radiology report generation for PET images. PET2Rep stands out as the first dedicated dataset for PET report generation with metabolic information, uniquely capturing whole-body image-report pairs that cover dozens of organs to fill the critical gap in existing benchmarks and mirror real-world clinical comprehensiveness. In addition to widely recognized natural language generation metrics, we introduce a series of clinical efficacy metrics to evaluate the quality of radiotracer uptake pattern description in key organs in generated reports. We conduct a head-to-head comparison of 30 cutting-edge general-purpose and medical-specialized VLMs. The results show that the current state-of-the-art VLMs perform poorly on PET report generation task, falling considerably short of fulfilling practical needs. Moreover, we identify several key insufficiency that need to be addressed to advance the development in medical applications. We believe PET2Rep will serve as a platform for the development and application of VLMs for PET imaging, accelerating the development of trustworthy reporting tools that can genuinely alleviate radiologist burden and enhance patient care. \textbf{Project page: }https://github.com/YichiZhang98/PET2Rep.

\end{abstract}

\begin{figure}[t]
	\includegraphics[width=\linewidth]{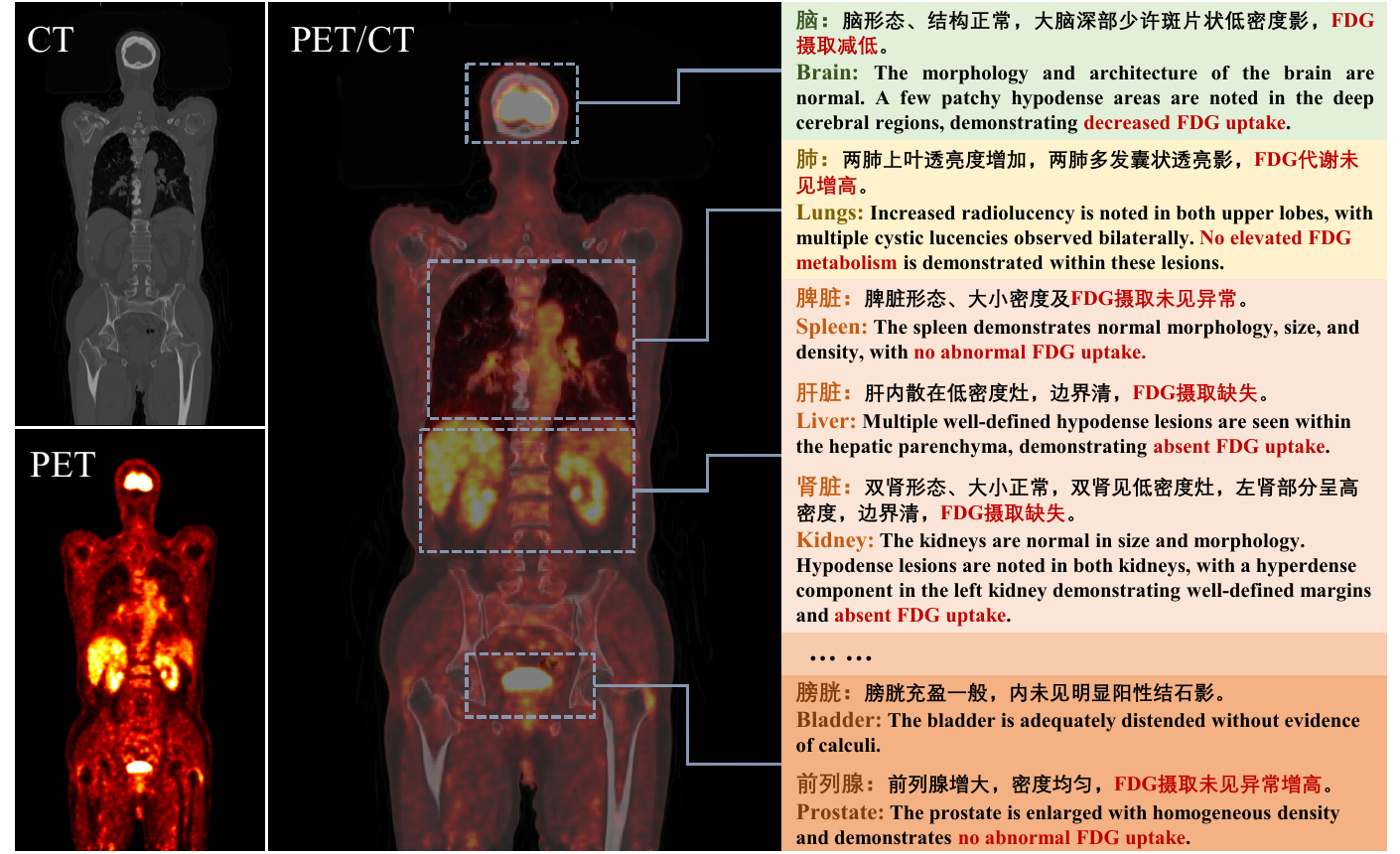}
	\caption{An overview of PET2Rep benchmark. Each case contains whole-body PET/CT images with radiology report. }
	\label{Abstract}
\end{figure}

\section{Introduction}

Radiology plays a crucial role in modern healthcare, enabling clinicians to visualize internal structures of patients and make informed decisions for diagnosis and treatment. 
Positron Emission Tomography (PET) stands as a cornerstone in contemporary oncological and neurological imaging, offering unparalleled insights into dynamic metabolic processes.
Unlike imaging modalities like X-ray and CT which primarily focus on information of anatomical structures, PET excels at visualizing metabolic information of physiological functions. By tracking the distribution of radioactive tracers, PET can detect early signs of disease progression, monitor treatment response, and guide personalized therapy plans \cite{peng202318f,xue202418f}. This functional imaging capability has revolutionized the diagnosis and management of various conditions \cite{schwenck2023advances}.
In the clinical workflow, radiology reports play a pivotal role in translating imaging into actionable information for healthcare providers. These reports summarize the radiologist's interpretation of the images, highlight key observations, and provide recommendations for further evaluation or treatment \cite{pang2023survey}. However, the process of manually creating these reports is inherently labor-intensive and time-consuming, often burdening radiologists with a significant administrative workload. This not only limits their capacity to handle a growing volume of imaging studies, but also introduces potential delays in patient care \cite{ashraf2023incidence}.

The recent surge in vision-language models (VLMs) has attracted interest from the medical community, where these models hold great potential to transform various aspects of clinical practice like automatic medical report generation \cite{zhang2024vision}. Leveraging the power of large-scale pretraining, VLMs can analyze medical images and generate corresponding textual descriptions, effectively bridging the gap between visual data and clinical language. However, existing applications of VLMs in the medical domain have predominantly focused on structural imaging modalities \cite{liu2024gemex,hamamci2024developing,zhu2025how}, while the unique characteristics and clinical value of PET imaging have largely been overlooked in the current research landscape.
As the analysis of PET images poses unique challenges due to the need to integrate functional and anatomical information and specialized knowledge required to interpret tracer uptake patterns \cite{coleman2010pet,matsubara2022review}, it is worth rethinking that \textit{How Far are VLMs from Effective Radiology Report Generation for Positron Emission Tomography Imaging?}

To answer this question, we introduce PET2Rep, a comprehensive benchmark for the evaluation of radiology report generation for PET imaging. 
Compared with existing medical benchmarks, the key advantages of PET2Rep can be concluded in the following three aspects.

\textbf{The First PET/CT Report Dataset.}
PET2Rep is the first dataset dedicated to PET/CT report generation. Unlike other modalities like X-ray and CT which primarily focus on anatomical structures, PET operates at the molecular level, enabling the assessment of metabolic information. This unique feature allows for early disease identification, often before anatomical changes are visible on other imaging modalities \cite{gatidis2024results}.
PET2Rep is a large-scale multi-modal dataset of 565 cases with paired PET, CT and corresponding radiology reports.
Given the high cost of PET/CT scans and the need for specialized expertise in report writing, there is currently no relevant dataset available, which highlights the importance of PET2Rep in advancing research in this field.

\textbf{Whole-Body Imaging with Radiology Reports.}
Existing medical imaging benchmarks are often limited to specific anatomical domains. For instance, chest X-ray report generation primarily address thoracic pathologies \cite{liu2024gemex}, while those for CT reports concentrate on the volume and morphology of organs and lesions in the chest \cite{hamamci2024developing} or abdominal regions \cite{bassi2025radgpt}. In contrast, PET2Rep encompasses a much broader anatomical scope, with images ranging from the head and neck to the proximal limbs. Consequently, its corresponding reports provide detailed evaluations of dozens of organs body-wide, demanding a more extensive scope of medical knowledge for accurate interpretation, as shown in Figure.\ref{Abstract}. This holistic approach more closely simulates real-world oncology practice, where radiologists conduct comprehensive assessments rather than focusing on isolated areas.

\textbf{Data Collection from Clinical Scenarios.}
Many existing medical multimodal benchmarks are developed from public imaging archives \cite{sepehri2024mediconfusion,huatuogpt}. These frameworks often generate tasks that probe for superficial understanding of the image, such as identifying the imaging modality or naming marked organs, rather than complex clinical reasoning \cite{ye2024gmai,zhou2025drvd}.
Such scenarios test for basic medical knowledge and differ significantly from the complex demands of a real clinical workflow. In contrast, PET2Rep is collected from real clinical scenarios and incorporates data directly from the clinical setting, ensuring that the benchmark authentically reflects the challenges radiologists encounter in their daily work.
This ensures the authenticity and clinical relevance of the PET2Rep benchmark while minimizing the risk of data leakage, thereby reflecting the generalization performance of VLMs in real-world clinical scenarios.

To make a comprehensive evaluation of the performance of VLMs, we establish a standardized evaluation pipeline for PET/CT radiology report generation.
We formulate a prompting framework incorporating essential elements including imaging modality specifications and clinical objectives and design a structured report template aligned with radiological training protocols. This approach ensures faithful translation of image-derived information into formatted reports that maintain consistency with expert-generated radiological reports.
We conduct a comprehensive evaluation state-of-the-art models, including 19 general purpose and 11 medical-specific VLMs on PET2Rep benchmark.
The experimental results show that current cutting-edge VLMs exhibit suboptimal performance on the task, falling considerably short of fulfilling real-world requirements. Furthermore, our analysis reveals several critical limitations that must be tackled to drive progress in clinical applications.

\begin{figure*}[t!]
	\includegraphics[width=\linewidth]{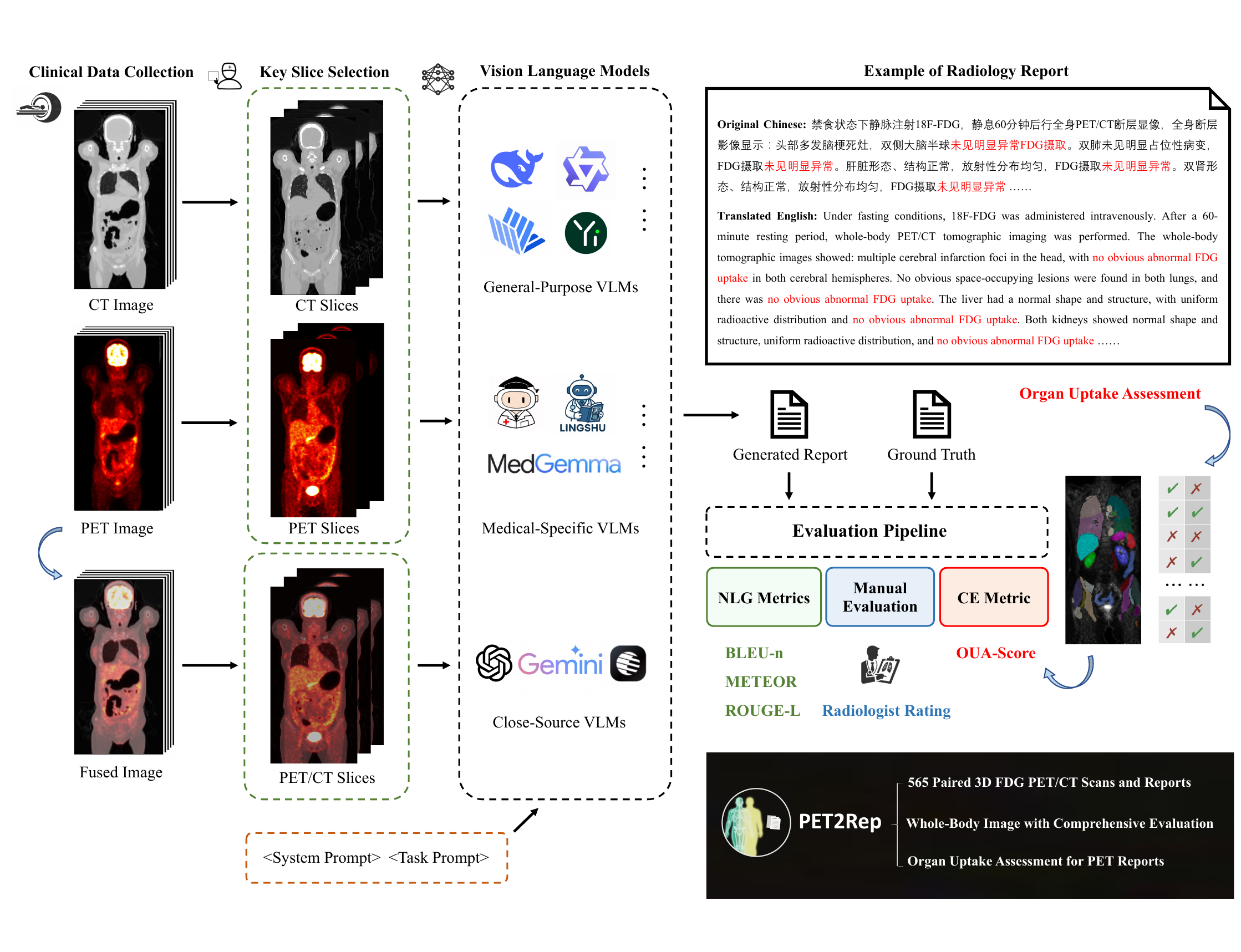}
	\caption{Pipeline of the PET2Rep benchmark for evaluation of VLM-based PET radiology report generation. First, PET/CT images are analyzed by VLMs with a designed prompt format to include necessary information such as image modality, clinical task, and designed report template based on radiologist training guidelines. Then the generated reports are evaluated against the ground-truth reports with widely recognized natural language generation (NLG) metrics and a novel clinical efficacy (CE) metric for PET imaging. We further conduct manual scoring by radiologists for more comprehensive evaluation.}
	\label{Benchmark}
\end{figure*}

\section{Related Works}

\subsection{Positron Emission Tomography}

Positron Emission Tomography (PET) is a clinical imaging technique that reveals ongoing metabolic processes in the body by detecting gamma photons generated from positron annihilation after injecting radioactive tracers. As the most widely used tracer, fluorodeoxyglucose (FDG) assesses local glucose uptake to evaluate organ metabolism and detect tumor metastasis, enabling monitoring of treatment progress \cite{ren2019atlas}.
Clinically, PET is primarily used for early tumor screening for cancer detection \cite{gatidis2024results,peng202318f}, organ metabolic function assessment \cite{xue202418f,zhang2025seganypet}, and treatment monitoring \cite{van2021diagnostic}.
The unique metabolic imaging capability provides indispensable insights for disease diagnosis and treatment optimization in clinical applications.

\subsection{Vision-Language Models}

Vision-Language Models (VLMs) have emerged as a transformative development in artificial intelligence, effectively bridging the gap between visual perception and natural language understanding \cite{zhang2024vision}. The swift progress in VLM development is largely attributed to innovative pre-training strategies and architectural designs, which have demonstrated remarkable capabilities across a wide array of tasks
such as visual question answering and image captioning \cite{chen2022visualgpt,feng2025vision,lin2025rs}.
Beyond general-purpose vision tasks, VLMs are making significant inroads into specialized fields like medical image analysis \cite{peng2025scaling}.
VLMs can generate diagnostic reports, answer clinical questions, and highlight regions of interest, offering substantial support to healthcare professionals and promising to enhance the efficacy and accuracy of medical diagnoses \cite{zhang2024segment,jiang2024evaluating,lin2025taming}.

\subsection{Radiology Report Generation}

The core task of radiology report generation is to transform medical imaging information into accurate and standardized textual reports. 
Early studies primarily focused on training encoder-decoder architectures for report generation, where the image features are extracted by an encoder and then fed into a decoder to predict the corresponding report \cite{jing2018automatic}.
Given the complexity and the inherent variability in radiological findings, several approaches utilize confidential guidance or attention mechanism to enhance the adaptability \cite{song2022cross,wang2024trust}.
Due to the impressive performance in a variety of downstream tasks \cite{zhang2024vision}, there has been a surge of investigating VLMs for radiology report generation \cite{hamamci2024ct2rep,chen2025large}.

\section{PET2Rep Benchmark}

We introduce PET2Rep, a comprehensive benchmark designed to evaluate the performance of VLMs for generating radiology reports from PET images.
PET2Rep is the first PET/CT dataset with paired structured radiology reports.
A key distinction of PET2Rep is its data sourcing. Unlike benchmarks that rely on data aggregated from online repositories, all data in our work were meticulously collected from actual clinical scenarios. This approach guarantees the authenticity and clinical relevance of the benchmark. Furthermore, by sourcing directly from clinical settings, we mitigate the risk of data leakage, ensuring that PET2Rep accurately reflects the complexity and diversity of real-world radiological practice. The comprehensive workflow of the PET2Rep benchmark is detailed in Figure. \ref{Benchmark} and will be elaborated upon in the subsequent sections.

\subsection{Dataset Construction}

We collect 565 cases of 3D whole-body FDG PET/CT imaging from one local medical center, which is the most widely used PET tracer in oncology.
As a non-specific tracer, FDG can be used for whole-body imaging to reflect tissue glucose metabolism, which makes the imaging useful in assessing the systemic distribution and metastasis of tumors.
Structured radiology reports are constructed based on radiologist-designed templates, which play a pivotal role in assisting physicians to interpret whole-body PET/CT scans in a standardized and organized manner, thereby enhancing clarity and supporting clinical decision-making. The report provides a detailed and objective description of the findings from the PET/CT images in a systematic, head-to-toe sequence, ensuring that no anatomical region is overlooked. It enumerates all detected abnormalities without offering interpretative conclusions, serving as a factual foundation for subsequent clinical assessment.
More detailed information and examples of the dataset are shown in the Appendix.

\subsection{Data Preprocessing}

To ensure the accuracy and reliability of multi-modal image analysis, rigorous data pre-processing is indispensable for bringing all imaging modalities into a consistent and interpretable format.
In PET/CT imaging, a critical pre-processing step involves resampling the CT images to match the lower spatial resolution of the PET images. This coregistration process aligns both modalities to a common matrix size, ensuring voxel-wise correspondence across datasets. Following resampling, the CT intensities are standardized using z-score normalization to reduce inter-scan variability. Additionally, normalization of PET data is performed by converting the raw radioactivity counts into Standardized Uptake Values (SUV) a widely adopted quantitative metric in PET imaging that accounts for factors such as the injected radiotracer dose and the patient’s body weight \cite{lucignani2004use}.
To emulate the clinical diagnostic workflow, we implement a fusion process that integrates PET and CT scans. This approach combines the functional information from PET with the anatomical detail provided by CT, reflecting the manner in which radiologists interpret these modalities in clinical practice. The resulting composite image enables visualization of metabolic activity within its precise anatomical context. Such integration is essential for accurately localizing regions of abnormal radiotracer uptake and facilitating a comprehensive assessment of the patient’s condition.

\subsection{Key Slice Selection}

Given that the original PET/CT images are three-dimensional, while most existing VLMs are designed for 2D images, it becomes necessary to select out representative 2D slices from the 3D imaging.
In this study, we select the coronal plane as the view for slice sampling, following clinical conventions in which radiologists commonly utilize this view for comprehensive head-to-toe assessments. 
The coronal plane offers an optimal perspective, capturing the global anatomical context and encompassing the majority of key organs. By analyzing multiple coronal slices, VLMs can effectively capture the salient information embedded within the full 3D scan.
Building upon this design, we further emulate the diagnostic process of radiologists and design two strategies for report generation as described below.

\textbf{Input Separate PET and CT Images.}
In this strategy, we maintain the distinction between functional and anatomical information by providing the model with two distinct, parallel inputs. For each anatomical location of interest, we extract a corresponding pair of 2D slices: a grayscale slice from the CT volume for structural context, and a pseudo-colored slice from the PET volume to highlight metabolic activity. Specifically, we identify three key locations for analysis, resulting in a total input of six images for the VLMs (a PET/CT pair for each location). This dual-input approach compels the model to learn the complex correlations between anatomical structure and functional uptake, mirroring the cognitive process of a radiologist integrating two different sets of images.

\textbf{Input Fused PET/CT Images.}
This strategy involves pre-integrating the multimodal information into a single image before presenting it to the model. For each selected location, we generate a fused image by superimposing the pseudo-colored PET slice directly onto its corresponding grayscale CT slice. In our implementation, we create these fused images for three key locations, providing the model with a total of three input PET/CT images. Each fused image presents an immediate composite view, in which metabolic hotspots are precisely localized within their anatomical context. This method simulates the final visualization that radiologists use for diagnosis. By supplying the model with pre-fused inputs, we eliminate the need for it to learn the fusion process, enabling it to focus directly on understanding the integrated functional and structural patterns.

\subsection{Experimental Setup}

In this study, we evaluate a range of VLMs encompassing both medical-specific and general-purpose models including open-source options and those accessible via proprietary APIs.
The weights of open-source models were sourced from respective official Hugging Face repositories. 
To guide the models in generating radiology-style reports, we design a standardized prompt format specifying the imaging modality, clinical task, and a report template derived from radiologist training guidelines. This ensures that image interpretations are expressed in a format consistent with manually authored radiological reports.
Our evaluation are conducted in a zero-shot setting, which serves as a stringent test of the models' generalization ability, revealing how well they can handle complex medical imaging tasks without any task-specific fine-tuning.
All tests were executed using NVIDIA A100 GPUs with 80GB of memory.

\subsection{Evaluated Models}

To comprehensively assess the performance of VLMs on the PET2Rep benchmark, we conducted a systematic evaluation of diverse state-of-the-art VLMs, spanning general-purpose models, medical-specific models, and closed-source models tested via API access. 

\textbf{General-Purpose VLMs.}
General-purpose VLMs are trained on large and diverse datasets to perform a wide spectrum of multimodal tasks. Their broad training enables strong visual understanding and reasoning capabilities, with versatility and scalability. We select following general-purpose VLMs for evaluation, including Qwen2.5-VL series \cite{Qwen25vl}, InternVL3 series \cite{Internvl3}, Yi-VL series \cite{Yi}, LLaVA-V1.5 \cite{Llava}, OmniLMM \cite{Omnilmm}, VisualGLM \cite{GLM} and Deepseek-VL2 \cite{Deepseekvl2} . 

\textbf{Medical-Specific VLMs.} 
In contrast to general-purpose models, medical-specific VLMs are tailored for clinical applications, emphasizing domain adaptation and integration of specialized medical knowledge. Trained on curated medical datasets and aligned with diagnostic workflows, these models prioritize accuracy and reliability in healthcare settings.
We select following medical-specific VLMs for evaluation, including LLaVA-Med \cite{Llavamed}, Med-Flamingo \cite{Medflamingo}, Qilin-Med-VL \cite{Qilin}, RadFM \cite{RadFM}, MedDr \cite{Meddr}, HuatuoGPT-Vision \cite{huatuogpt}, MedVLM-R1 \cite{Medvlmr1}, and latest MedGemma series \cite{medgemma} and Lingshu series \cite{Lingshu}.

\textbf{Closed-Source VLMs.}
Closed-source VLMs are developed and maintained by enterprises with inaccessible source code, typically provided to users via APIs for integration into applications.
We select following closed-source VLMs for evaluation, including Gemini 2.5 Pro \cite{gemini25}, GPT-4o \cite{GPT-4o}, Moonshot-v1 \cite{moonshot} and Qwen-VL-Max \cite{Qwenvl}.

\subsection{Evaluation Metrics}

To assess the performance of VLMs in radiology report generation, we compare the generated reports against the ground-truth reports using the following aspects.

\textbf{Natural Language Generation (NLG) Metrics.}
In line with existing research, we adopt widely recognized NLG metrics, including BLEU-n \cite{bleu}, METEOR \cite{meteor}, and ROUGE-L \cite{rouge}. 
Specifically, BLEU‑n evaluates n‑gram overlap between generated and reference reports, ROUGE‑L measures alignment via the longest common subsequences, and METEOR accounts for synonyms and paraphrases to capture semantic similarity.

\textbf{PET Clinical Efficacy (CE) Metrics.}
NLG metrics primarily focus on word and sentence similarity while neglecting diagnostic accuracy. 
Reports with opposite diagnostic conclusions may achieve similar NLG scores. Conversely, results with correct uptake assessments but inconsistent formatting in the report text might receive lower NLG scores. Existing studies have explored the proposal of clinical efficacy metrics by utilizing text classifiers to extract abnormality labels for CT report evaluation \cite{hamamci2024developing}. However, these methods are not applicable to PET reports.
To assess the clinical efficacy of PET reports, we introduce a series of CE metrics to evaluate descriptions regarding radiotracer uptake patterns in key organs within generated PET reports.
Given that the whole-body PET imaging data used in our study, we extract the assessment of uptake levels corresponding to each key organ from the report text and compare these assessments with the corresponding entries in the ground truth reports.
For each key organ, we define four states of radiotracer uptake: \textit{Increased Uptake}, \textit{Decreased Uptake}, \textit{Absent Uptake}, and \textit{Normal}. 
Given the clinical focus on anomaly detection, we categorize the first three states into three distinct positive classes, with \textit{Normal} serving as the negative class.
Our evaluation method involves independently calculating the precision, recall, and F1-score for each of the three positive classes.
The final CE metrics are the macro-average across these three positive classes.
The implementation details are elaborated in the Appendix.
Compared to NLG metrics, CE metrics shift the evaluation from text-matching problem to multi-label classification assessment that more closely aligns with clinical diagnosis.

\begin{table*}[!htbp]
\centering
\footnotesize
\setlength\tabcolsep{3pt}
\renewcommand\arraystretch{1.15}
\begin{tabular}{l|ccc|ccc|c}
\hline \hline
\multirow{2}{*}{\textbf{Model} (year/month)} & \multicolumn{3}{c|}{\textbf{NLG Metrics}} & \multicolumn{3}{c|}{\textbf{CE Metrics}} & \textbf{Overall} \\ 
\cline{2-7} & \textbf{BL-4} & \textbf{MTR} & \textbf{RG-L} & \textbf{Pre} & \textbf{Rec} & \textbf{F1} & \textbf{(\%)} \\ \hline \hline
Template Baseline & 0.3150(0.0482) & 0.1475(0.0141) & 0.5110(0.0319) & 0.2282(0.0179) & 0.2220(0.0106) & 0.2249(0.0123) & 27.5 \\ \hline
\multicolumn{8}{c}{\textbf{General-Purpose VLMs}} \\ \hline
\multirow{2}{*}{Qwen2.5-VL-7B (25/1)} & \cellcolor{gray!15}0.3050(0.0476) & \cellcolor{gray!15}0.1407(0.0198) & \cellcolor{gray!15}0.5075(0.0340) & \cellcolor{gray!15}0.2233(0.0236) & \cellcolor{gray!15}0.1974(0.0083) & \cellcolor{gray!15}0.2094(0.0132) & \cellcolor{gray!15}26.4 \\
& 0.3057(0.0467) & 0.1390(0.0186) & 0.5088(0.0320) & 0.2284(0.0227) & 0.2023(0.0075) & 0.2144(0.0121) & 26.6 \\ \hline
\multirow{2}{*}{Qwen2.5-VL-32B (25/1)} & \cellcolor{gray!15}0.1777(0.0421) & \cellcolor{gray!15}0.0063(0.0110) & \cellcolor{gray!15}0.4165(0.0516) & \cellcolor{gray!15}0.3402(0.0781) & \cellcolor{gray!15}0.0418(0.0127) & \cellcolor{gray!15}0.0743(0.0214) & \cellcolor{gray!15}17.6 \\
& 0.1851(0.0408) & 0.0063(0.0111) & 0.4295(0.0486) & 0.2728(0.0447) & 0.0308(0.0047) & 0.0554(0.0082) & 16.3 \\ \hline
\multirow{2}{*}{Qwen2.5-VL-72B (25/1)} & \cellcolor{gray!15}0.2223(0.0585) & \cellcolor{gray!15}0.0655(0.0172) & \cellcolor{gray!15}0.4234(0.0588) & \cellcolor{gray!15}0.2474(0.0513) & \cellcolor{gray!15}0.0295(0.0024) & \cellcolor{gray!15}0.0527(0.0043) & \cellcolor{gray!15}17.3 \\
& 0.2273(0.0584) & 0.0645(0.0171) & 0.4306(0.0594) & 0.2917(0.0328) & 0.0393(0.0049) & 0.0693(0.0084) & 18.7 \\ \hline
\multirow{2}{*}{InternVL3-8B (25/4)} & \cellcolor{gray!15}0.2439(0.0627) & \cellcolor{gray!15}0.0606(0.0443) & \cellcolor{gray!15}0.4739(0.0630) & \cellcolor{gray!15}0.2425(0.0151) & \cellcolor{gray!15}0.2107(0.0114) & \cellcolor{gray!15}0.2254(0.0119) & \cellcolor{gray!15}24.3 \\
& 0.2509(0.0529) & 0.0641(0.0463) & 0.4845(0.0566) & 0.2333(0.0153) & 0.2099(0.0074) & 0.2208(0.0087) & 24.4 \\ \hline
\multirow{2}{*}{InternVL3-14B (25/4)} & \cellcolor{gray!15}0.2513(0.0684) & \cellcolor{gray!15}0.0472(0.0528) & \cellcolor{gray!15}0.4835(0.0910) & \cellcolor{gray!15}0.2366(0.0206) & \cellcolor{gray!15}0.2057(0.0095) & \cellcolor{gray!15}0.2199(0.0129) & \cellcolor{gray!15}24.1 \\
& 0.2495(0.0671) & 0.0532(0.0506) & 0.4813(0.0904) & 0.2322(0.0196) & 0.1982(0.0099) & 0.2137(0.0131) & 23.8 \\ \hline
\multirow{2}{*}{InternVL3-38B (25/4)} & \cellcolor{gray!15}0.1377(0.0924) & \cellcolor{gray!15}0.0775(0.0483) & \cellcolor{gray!15}0.4371(0.1199) & \cellcolor{gray!15}0.2711(0.0203) & \cellcolor{gray!15}0.2072(0.0141) & \cellcolor{gray!15}0.2344(0.0127) & \cellcolor{gray!15}22.8 \\
& 0.1446(0.0855) & 0.0825(0.0480) & 0.4618(0.0913) & 0.2674(0.0258) & 0.2435(0.0298) & 0.2546(0.0278) & 24.2 \\ \hline
\multirow{2}{*}{InternVL3-78B (25/4)} & \cellcolor{gray!15}0.3090(0.0525) & \cellcolor{gray!15}0.1233(0.0359) & \cellcolor{gray!15}0.4997(0.0401) & \cellcolor{gray!15}0.2355(0.0255) & \cellcolor{gray!15}0.0520(0.0119) & \cellcolor{gray!15}0.0850(0.0157) & \cellcolor{gray!15}21.7 \\
& 0.3090(0.0518) & 0.1262(0.0318) & 0.5008(0.0397) & 0.2369(0.0492) & 0.0748(0.0083) & 0.1132(0.0122) & 22.7 \\ \hline
\multirow{2}{*}{Yi-VL-6B (24/1)} & \cellcolor{gray!15}0.0065(0.0316) & \cellcolor{gray!15}0.0002(0.0056) & \cellcolor{gray!15}0.0479(0.0709) & \cellcolor{gray!15}0.1144(0.0430) & \cellcolor{gray!15}0.0061(0.0020) & \cellcolor{gray!15}0.0115(0.0038) & \cellcolor{gray!15}3.1 \\
& 0.0374(0.0733) & 0.0029(0.0165) & 0.1156(0.1432) & 0.1519(0.0261) & 0.0260(0.0033) & 0.0444(0.0055) & 6.3 \\ \hline
\multirow{2}{*}{Yi-VL-34B (24/1)} & \cellcolor{gray!15}0.2610(0.1071) & \cellcolor{gray!15}0.0848(0.0664) & \cellcolor{gray!15}0.4439(0.1420) & \cellcolor{gray!15}0.2305(0.0159) & \cellcolor{gray!15}0.1869(0.0079) & \cellcolor{gray!15}0.2063(0.0098) & \cellcolor{gray!15}23.6 \\
& 0.2854(0.0809) & 0.0898(0.0645) & 0.4779(0.0950) & 0.2303(0.0211) & 0.2038(0.0072) & 0.2160(0.0116) & 25.1 \\ \hline
\multirow{2}{*}{LLaVa-V1.5-7B (23/9)} & \cellcolor{gray!15}0.1198(0.0508) & \cellcolor{gray!15}0.0126(0.0515) & \cellcolor{gray!15}0.3043(0.0639) & \cellcolor{gray!15}0.2044(0.0287) & \cellcolor{gray!15}0.1022(0.0091) & \cellcolor{gray!15}0.1306(0.0121) & \cellcolor{gray!15}14.6 \\
& 0.0328(0.0141) & 0.0056(0.0369) & 0.1717(0.0283) & 0.2460(0.0764) & 0.0337(0.0093) & 0.0592(0.0163) & 9.2 \\ \hline
\multirow{2}{*}{OmniLMM-12B(24/4)} & \cellcolor{gray!15}0.0412(0.0627) & \cellcolor{gray!15}0.0075(0.0232) & \cellcolor{gray!15}0.1339(0.1324) & \cellcolor{gray!15}0.1789(0.0330) & \cellcolor{gray!15}0.0173(0.0027) & \cellcolor{gray!15}0.0316(0.0050) & \cellcolor{gray!15}6.8 \\
& 0.0397(0.0614) & 0.0067(0.0238) & 0.1293(0.1336) & 0.2095(0.0393) & 0.0180(0.0040) & 0.0331(0.0071) & 7.3 \\ \hline
\multirow{2}{*}{VisualGLM-6B 23/5} & \cellcolor{gray!15}0.0361(0.0519) & \cellcolor{gray!15}0.0182(0.0517) & \cellcolor{gray!15}0.1338(0.1214) & \cellcolor{gray!15}0.0662(0.0710) & \cellcolor{gray!15}0.0002(0.0002) & \cellcolor{gray!15}0.0004(0.0004) & \cellcolor{gray!15}4.3 \\
& 0.0306(0.0492) & 0.0208(0.0588) & 0.1173(0.1157) & 0.3404(0.1494) & 0.0014(0.0006) & 0.0029(0.0012) & 8.6 \\ \hline
\multirow{2}{*}{DeepSeek-VL2 (24/12)} & \cellcolor{gray!15}0.2697(0.0675) & \cellcolor{gray!15}0.0939(0.0976) & \cellcolor{gray!15}0.4875(0.0536) & \cellcolor{gray!15}0.2170(0.0137) & \cellcolor{gray!15}0.1532(0.0076) & \cellcolor{gray!15}0.1795(0.0081) & \cellcolor{gray!15}23.4 \\
& 0.2817(0.0637) & 0.1054(0.0974) & 0.4936(0.0476) & 0.2198(0.0269) & 0.1571(0.0135) & 0.1831(0.0176) & 24.0 \\ \hline
\multicolumn{8}{c}{\textbf{Medical-Specific VLMs}} \\ \hline
\multirow{2}{*}{MedDr(24/4)} & \cellcolor{gray!15}0.2667(0.1012) & \cellcolor{gray!15}0.1564(0.0434) & \cellcolor{gray!15}0.4571(0.1168) & \cellcolor{gray!15}0.2270(0.0245) & \cellcolor{gray!15}0.1820(0.0201) & \cellcolor{gray!15}0.2020(0.0215) & \cellcolor{gray!15}24.9 \\
& 0.2801(0.0874) & 0.1536(0.0389) & 0.4742(0.0951) & 0.2397(0.0275) & 0.2113(0.0084) & 0.2243(0.0138) & 26.4 \\ \hline
\multirow{2}{*}{HuatuoGPT-Vision (24/6)} & \cellcolor{gray!15}0.1384(0.0865) & \cellcolor{gray!15}0.0000(0.0000) & \cellcolor{gray!15}0.3399(0.1112) & \cellcolor{gray!15}0.1692(0.0232) & \cellcolor{gray!15}0.0814(0.0186) & \cellcolor{gray!15}0.1097(0.0207) & \cellcolor{gray!15}14.0 \\
& 0.2573(0.0546) & 0.0743(0.0278) & 0.4834(0.0577) & 0.2183(0.0200) & 0.1620(0.0148) & 0.1859(0.0164) & 23.0 \\ \hline
\multirow{2}{*}{MedVLM-R1 (25/2)} & \cellcolor{gray!15}0.1602(0.1112) & \cellcolor{gray!15}0.0006(0.0097) & \cellcolor{gray!15}0.3472(0.1742) & \cellcolor{gray!15}0.2246(0.0285) & \cellcolor{gray!15}0.1019(0.0117) & \cellcolor{gray!15}0.1399(0.0150) & \cellcolor{gray!15}16.2 \\
& 0.1708(0.1294) & 0.0003(0.0070) & 0.3358(0.1840) & 0.2321(0.0324) & 0.1204(0.0077) & 0.1583(0.0110) & 17.0 \\ \hline
\multirow{2}{*}{MedGemma-4B (25/7)} & \cellcolor{gray!15}0.3015(0.0517) & \cellcolor{gray!15}0.1215(0.0384) & \cellcolor{gray!15}0.5077(0.0385) & \cellcolor{gray!15}0.2276(0.0185) & \cellcolor{gray!15}0.2260(0.0113) & \cellcolor{gray!15}0.2266(0.0129) & \cellcolor{gray!15}26.8 \\
& 0.2874(0.0773) & 0.1207(0.0339) & 0.4875(0.0786) & 0.2362(0.0162) & 0.2245(0.0091) & 0.2301(0.0103) & 26.4 \\ \hline
\multirow{2}{*}{MedGemma-27B (25/7)} & \cellcolor{gray!15}0.2185(0.0552) & \cellcolor{gray!15}0.0297(0.0157) & \cellcolor{gray!15}0.4390(0.0696) & \cellcolor{gray!15}0.2300(0.0375) & \cellcolor{gray!15}0.0391(0.0079) & \cellcolor{gray!15}0.0667(0.0130) & \cellcolor{gray!15}17.1 \\
& 0.2251(0.0574) & 0.0309(0.0153) & 0.4521(0.0781) & 0.2853(0.0435) & 0.0846(0.0141) & 0.1304(0.0201) & 20.1 \\ \hline
\multirow{2}{*}{Lingshu-7B (25/6)} & \cellcolor{gray!15}0.2848(0.0855) & \cellcolor{gray!15}0.1079(0.0727) & \cellcolor{gray!15}0.4793(0.0933) & \cellcolor{gray!15}0.2281(0.0162) & \cellcolor{gray!15}0.1970(0.0100) & \cellcolor{gray!15}0.2112(0.0097) & \cellcolor{gray!15}25.1 \\
& 0.2775(0.0945) & 0.1030(0.0748) & 0.4700(0.1119) & 0.2273(0.0220) & 0.1942(0.0106) & \cellcolor{gray!15}0.2093(0.0138) & 24.7 \\ \hline 
\multirow{2}{*}{Lingshu-32B (25/6)} & \cellcolor{gray!15}0.3050(0.0650) & \cellcolor{gray!15}0.1554(0.0520) & \cellcolor{gray!15}0.4999(0.0604) & \cellcolor{gray!15}0.2250(0.0178) & \cellcolor{gray!15}0.2035(0.0079) & \cellcolor{gray!15}0.2135(0.0109) & \cellcolor{gray!15}26.7 \\
& 0.2987(0.0713) & 0.1531(0.0597) & 0.4939(0.0733) & 0.2328(0.0151) & 0.2071(0.0071) & 0.2191(0.0091) & 26.8 \\ \hline 
\multicolumn{8}{c}{\textbf{Closed-Source VLMs}} \\ \hline
\multirow{2}{*}{Gemini 2.5 Pro (25/6)}  & \cellcolor{gray!15}0.1535(0.0411) & \cellcolor{gray!15}0.0186(0.0199) & \cellcolor{gray!15}0.3987(0.0438) & \cellcolor{gray!15}0.1705(0.0299) & \cellcolor{gray!15}0.0215(0.0056) & \cellcolor{gray!15}0.0381(0.0092) & \cellcolor{gray!15}13.4 \\
& 0.1536(0.0420) & 0.0199(0.0201) & 0.4025(0.0477) & 0.2394(0.0571) & 0.0311(0.0066) & 0.0550(0.0115) & 15.0 \\ \hline
\multirow{2}{*}{GPT-4o (24/5)}  & \cellcolor{gray!15}0.2023(0.0422) & \cellcolor{gray!15}0.0287(0.0160) & \cellcolor{gray!15}0.4023(0.0421) & \cellcolor{gray!15}0.3375(0.0891) & \cellcolor{gray!15}0.0527(0.0110) & \cellcolor{gray!15}0.0910(0.0185) & \cellcolor{gray!15}18.6 \\
& 0.2134(0.0425) & 0.0318(0.0132) & 0.4168(0.0412) & 0.2540(0.0450) & 0.0728(0.0085) & 0.1130(0.0135) & 18.5 \\ \hline 
\multirow{2}{*}{Moonshot-v1 (25/1)} & \cellcolor{gray!15}0.3064(0.0496) & \cellcolor{gray!15}0.1261(0.0301) & \cellcolor{gray!15}0.5157(0.0339) & \cellcolor{gray!15}0.2603(0.0220) & \cellcolor{gray!15}0.1457(0.0096) & \cellcolor{gray!15}0.1866(0.0117) & \cellcolor{gray!15}25.7 \\
& 0.2923(0.0464) & 0.1055(0.0302) & 0.5142(0.0316) & 0.2327(0.0232) & 0.1803(0.0132) & 0.2030(0.0160) & 25.5 \\ \hline 
\multirow{2}{*}{Qwen-VL-Max (25/1)}  & \cellcolor{gray!15}0.2315(0.0375) & \cellcolor{gray!15}0.0269(0.0035) & \cellcolor{gray!15}0.4462(0.0377) & \cellcolor{gray!15}0.2764(0.0313) & \cellcolor{gray!15}0.1897(0.0139) & \cellcolor{gray!15}0.2248(0.0183) & \cellcolor{gray!15}23.3 \\
& 0.2479(0.0406) & 0.0265(0.0047) & 0.4649(0.0409) & 0.2844(0.0220) & 0.1802(0.0092) & 0.2204(0.0113) & 23.7\\ \hline
\hline
\end{tabular}
\caption{Evaluation of general-purpose and medical-specific VLMs on PET2Rep benchmark. Evaluation results presented in gray and white represent the results of separate PET and CT images and fused PET/CT images, respectively.} 
\label{Table_specialist}
\end{table*}

\begin{figure}[t]
	\includegraphics[width=\linewidth]{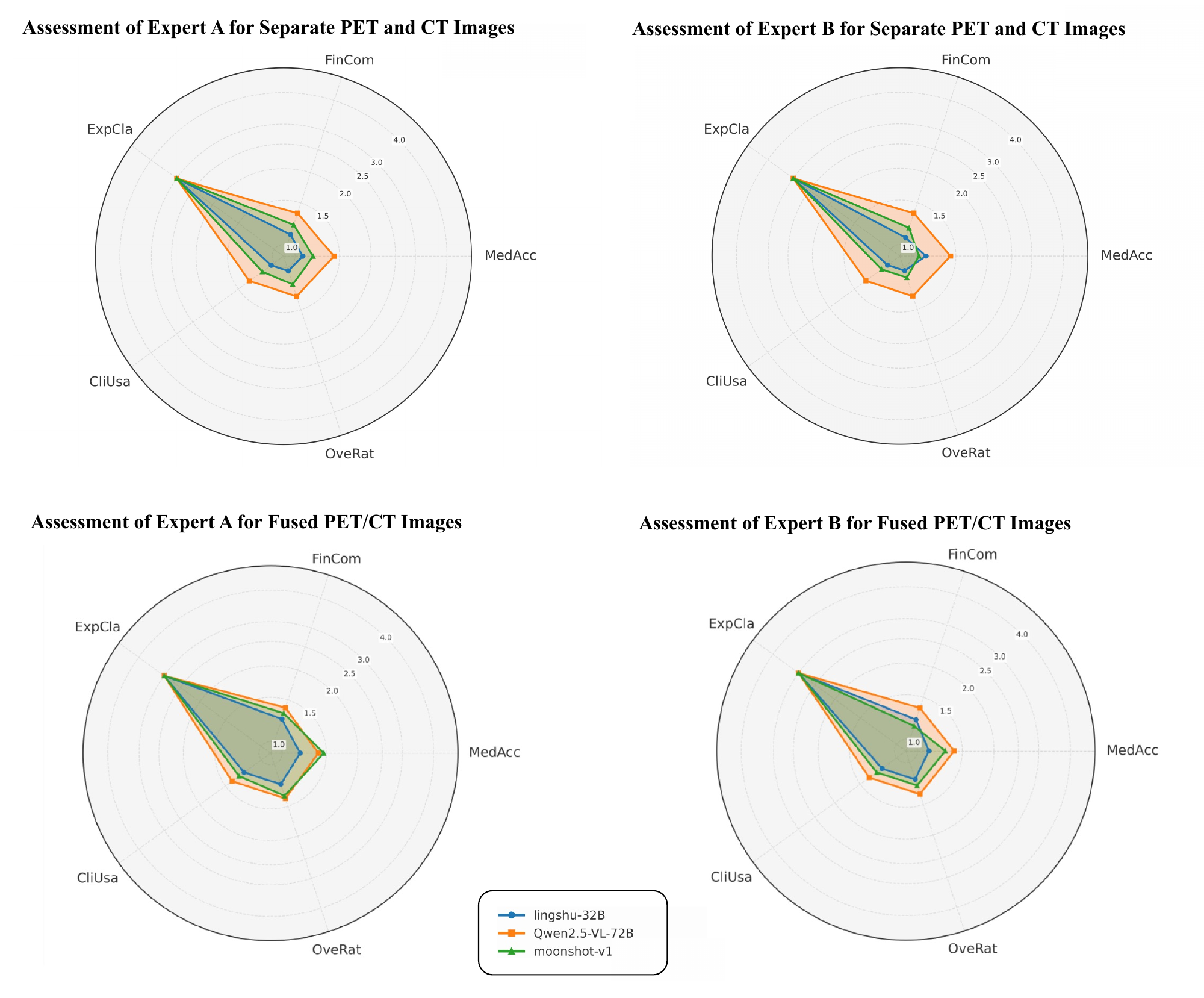}
	\caption{Performance comparison of three VLMs under different task settings for manual evaluation by two radiologists rated across five dimensions, including Medical Accuracy (MedAcc), Key Findings Completeness (FinCom), Expression Clarity (ExpCla), Clinical Usability (CliUsa) and Overall Rating (OveRat).}
	\label{Score}
\end{figure}

\section{Results and Analysis}

Table. \ref{Table_specialist} summarizes the performance of the evaluated VLMs, with the last column providing an overall reference score as the average of all metrics.
After reviewing the evaluation results, we have drawn following conclusions.

\textbf{Overall Ineffectiveness or Even Failure.}
All evaluated VLMs exhibit limited effectiveness in generating structured radiology reports. Alarmingly, most models fail to surpass even a simple template baseline. The requirement to produce comprehensive, whole-body structured reports presents a substantial challenge for existing VLMs. Many models are unable to consistently adhere to the prescribed report structure, occasionally generating disorganized, unusable, or even empty outputs, which yield near-zero scores across evaluation metrics and are thus omitted from the results table. Even when models attempt to follow the template, they often include irrelevant information or omit mandatory sections, underscoring their difficulty in capturing the core task requirements. This pattern suggests that many VLMs are overfitted to narrow training distributions, typically focused on specific tasks such as single-organ interpretation or generic image captioning, rather than holistic clinical reporting.
As a result, they struggle to generalize to clinical applications, where accuracy, completeness, and structural consistency are essential.

\textbf{State-of-the-Art Models Merely Match the Baseline.}
Although the most advanced models, such as the Lingshu and MedGemma series, outperform other VLMs, their performance remains only marginally comparable to the baseline. This underwhelming result indicates that even state-of-the-art VLMs are far from ready for practical application in clinical workflow.
While these models can generate coherent text with high NLG metrics, they frequently omit critical clinical details, such as subtle tracer uptake abnormalities, leading to low CE metrics.
Manual review by radiologists further confirms that the outputs of these models are largely unusable. The accurate interpretation of tracer uptake patterns, combined with the extensive medical knowledge required for comprehensive whole-body assessment, remains a major challenge, highlighting the gap between general language proficiency and specialized clinical expertise.
Further manual evaluation by two radiologists in Figure. \ref{Score} demonstrate that the outputs of state-of-the-art models are also mostly unusable.
The nuanced interpretation of tracer uptake patterns and the broad medical knowledge required for whole-body assessment remain significant challenges, highlighting a critical gap between general language proficiency and specialized clinical expertise.

\textbf{Larger Model Does Not Necessarily Translate to Better Performance.}
Our evaluation reveals an intriguing phenomenon that within the same model series, larger-scale models do not consistently outperform their smaller counterparts. In some cases, larger models appear to overlook task requirements, generating irrelevant or fabricated details such as patient names and ages, which negatively impact their evaluation performance. This observation suggests that the inferior performance of larger models may not stem from model scaling itself, but rather from insufficient exposure to domain-specific data and task-oriented training. Therefore, for specialized and highly structured tasks like PET report generation, architectural innovation and targeted fine-tuning may play a more critical role.

Further details regarding experimental results analysis and case studies are presented in the Appendix.

\section{Discussion and Conclusion}

In this work, we present PET2Rep, the first comprehensive benchmark specifically designed for evaluating radiology report generation in PET imaging, addressing a critical gap between existing research and clinical application.
The benchmark consists of 565 whole-body PET/CT image-report cases, representing a significant advancement in this domain. Another key innovation is the introduction of a series of clinical efficacy metrics to evaluate the quality of radiotracer uptake pattern description in key organs in generated reports PET reports, which is a decisive factor in clinical decision-making, as omissions in critical findings can alter therapeutic pathways.

Our experimental results clearly reveal the critical limitations of current VLMs. Despite their reported success on various multimodal medical benchmarks, all models fail to surpass even a simple template baseline in PET2Rep, with some models generating disorganized or structurally non-compliant reports.
These findings underscore the need for fundamental advancements in clinically grounded evaluation frameworks and rigorous alignment with real-world reporting standards to achieve genuine clinical applicability.
Many existing benchmark tasks assess the capabilities of VLMs through visual question answering, which primarily reflects superficial image understanding and falls short of the deep clinical reasoning required for diagnosis and treatment. Moreover, existing clinical report generation datasets are largely confined to localized anatomical structures, overlooking the integration of structural and functional information necessary for comprehensive whole-body evaluation.
In this context, PET2Rep serves as an expert-informed and clinically aligned benchmark that helps bridge this gap, providing a foundation for exploring the potential of large models toward more generalizable medical intelligence and facilitating progress in domain-specific model development.

While PET2Rep represents a significant step forward, several limitations should be acknowledged. At present, our evaluations are limited to 2D slices, which do not fully capture the three-dimensional spatial relationships and volumetric information critical for comprehensive image interpretation \cite{zhang2022bridging}. Moreover, clinically important quantitative indicators, such as standardized uptake values (SUVs) and lesion volume measurements, are not yet incorporated into the current evaluation framework.
To address these limitations, we plan to expand the benchmark to support full 3D PET/CT evaluations, enabling more complete spatial and volumetric analysis \cite{xue2025petwb}. Key quantitative measures, including SUVs and lesion volumes, will be reintegrated to enhance the benchmark’s clinical validity. In addition, while the current version supports only Chinese reports, future iterations will extend to multilingual evaluation, improving generalizability and facilitating broader clinical adoption of VLMs across diverse healthcare systems \cite{qiu2024towards}.
Another consideration lies in the contextual limitations of current VLMs. Generating long, structured whole-body reports may exceed the effective context length or reasoning capacity of some models.
In future work, we plan to further investigate this issue by exploring hierarchical strategies, such as summarizing findings by anatomical regions followed by integration into a standardized report template.
These planned enhancements will significantly improve the benchmark's clinical relevance and utility for developing more robust report generation systems.

\section{Acknowledgment}
{This work was supported by the National Natural Science Foundation of China (Grant No. 82394432 and 92249302), and the Shanghai Municipal Science and Technology Major Project (Grant No. 2023SHZDZX02). The computations in this research were performed using the CFFF platform of Fudan University.}

\bibliography{aaai26}

\clearpage

\section*{Appendix}

\subsection{A. Dataset Details}

\subsubsection{Data Collection.} All patients fasted for at least six hours and had blood glucose levels below 11.1 mmol/L before the scan. Patients were intravenously administered [18F]FDG at a dose of 3.70–5.55 MBq/Kg and then rested for a 60-minute uptake period.
Images were acquired using a Biograph 64 PET/CT scanner. The protocol included an initial CT scan (120 kV, 170 mA, 3.0 mm slice thickness) for attenuation correction, followed by a 3D PET scan over 5–6 bed positions with an acquisition time of 2.5 minutes per bed. Delayed imaging was performed in select cases. Final PET images were reconstructed using an iterative algorithm with CT-based attenuation correction.

\subsubsection{Quality Control.}
All images were visually inspected by two experienced nuclear medicine physicians. Additionally, cases with severe artifacts such as motion, truncation, and metal artifacts are excluded.

\subsubsection{Data Visualization.}
\begin{figure*}
	\includegraphics[width=\linewidth]{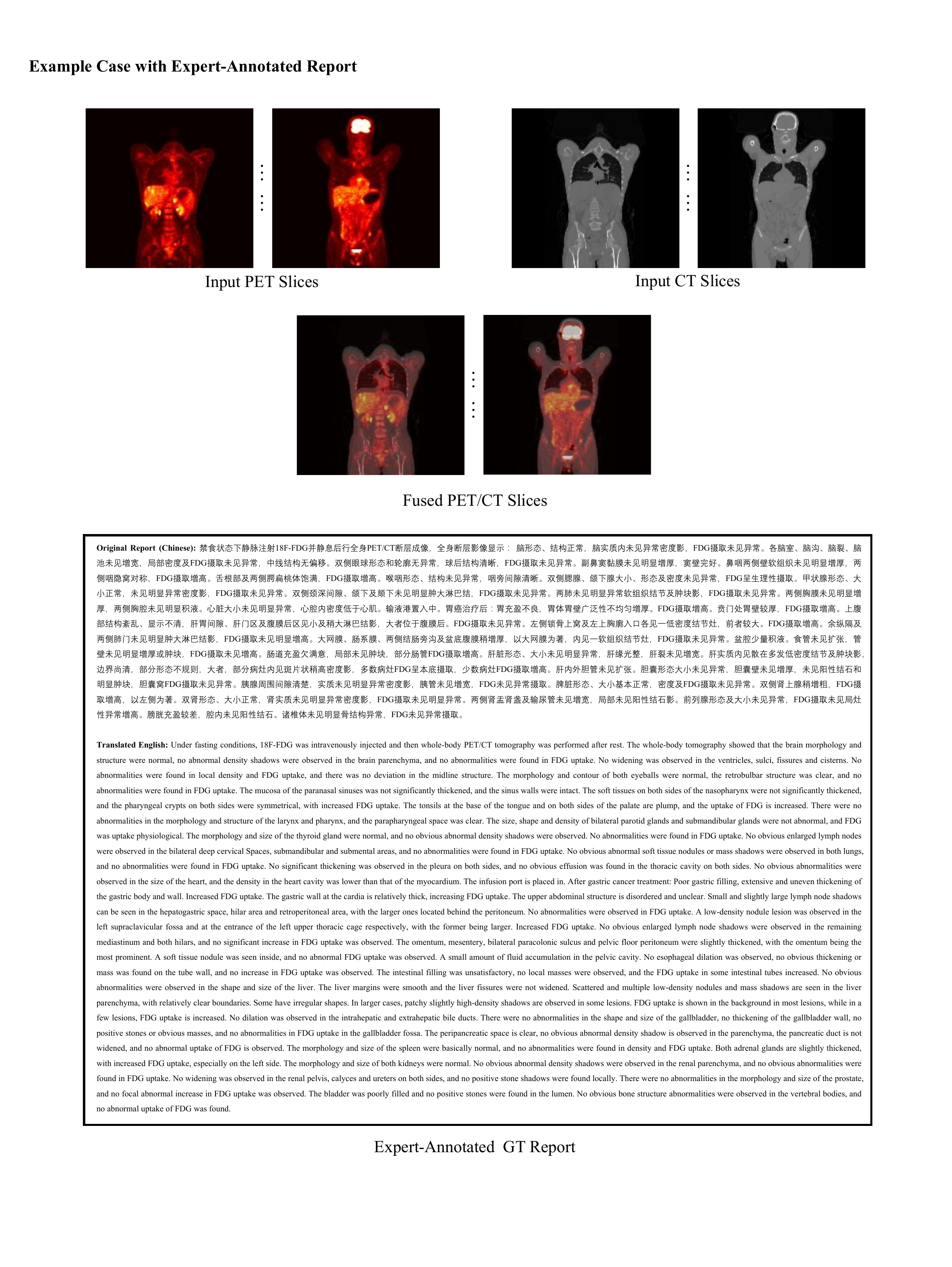}
	\caption{An example case with expert-annotated radiology report. }
	\label{Dataset}
\end{figure*}

Visualization example cases of PET2Rep benchmark are shown in Figure. \ref{Dataset}.

\subsection{B. Evaluation Details}

This appendix provides a detailed description of the implementation details of the evaluation metrics used for generated PET reports in our work.

\subsubsection{Natural Language Generation Metrics.}
To quantitatively assess the quality of generated Chinese reports, we developed a dedicated evaluation pipeline tailored to the linguistic idiosyncrasies of the language.
Initially, a rigorous preprocessing step is applied to both the generated and reference texts. This involves filtering the content to retain only Chinese characters, alphanumeric characters, and essential Chinese punctuation. Irrelevant special symbols are discarded to minimize noise while preserving semantic integrity. Given the absence of explicit word delimiters in written Chinese, we employ the Jieba segmentation library, a standard tool for Chinese word tokenization, to partition the character sequences into meaningful tokens. Any resulting empty or whitespace-only tokens are subsequently removed to ensure the fidelity of downstream calculations.
For metric computation, we adapted standard n-gram-based metrics. The BLEU-1 through BLEU-4 scores are calculated based on the overlap of n-grams in the segmented token sequences. We utilize uniform weights for individual BLEU scores and incorporate a smoothing function to mitigate the impact of data sparsity, a common issue in shorter texts with limited token overlap. To evaluate structural correspondence, we compute the ROUGE-L score by applying the longest common subsequence to the tokenized outputs. This approach effectively captures the sequential and contextual alignment between the generated and reference reports, which is crucial for evaluating narrative coherence in Chinese.

\begin{table}[t]
\centering
\setlength\tabcolsep{10pt}
\begin{tabular}{cl}
\hline
\textbf{No.} & \textbf{Organ (Anatomic Structures)} \\
\hline
1 & Cranium and Brain \\
2 & Eyeballs \\
3 & Nasal Cavity and Sinuses \\
4 & Pharynx and Parapharyngeal Space \\
5 & Palatine Tonsils and Larynx \\
6 & Salivary Glands and Thyroid \\
7 & Cervical Lymph Nodes \\
8 & Lungs and Thoracic Cavity \\
9 & Mediastinum and Heart \\
10 & Esophagus \\
11 & Liver \\
12 & Gallbladder \\
13 & Pancreas \\
14 & Spleen \\
15 & Kidneys and Adrenal Glands \\
16 & Gastrointestinal Tract \\
17 & Prostate/Uterus and Bladder \\
18 & Abdominal and Pelvic Cavities \\
19 & Spine and Bones \\
\hline
\end{tabular}
\caption{List of key organs (anatomic structures) evaluated in PET Report clinical efficacy metrics.}
\label{organ_list}
\end{table}

\subsubsection{PET Clinical Efficacy Metrics.}

Given that the whole-body PET imaging data used in our study, we extract the assessment of uptake levels corresponding to each key organ from the report text and compare these assessments with the corresponding entries in the ground truth reports. Our evaluation focuses on the assessment of radiotracer uptake for 19 predefined key organs and structures shown in Table. \ref{organ_list}.
For each key organ, we extract its uptake status from the report text and classify it into one of four mutually exclusive states of radiotracer uptake: \textit{Increased Uptake}, \textit{Decreased Uptake}, \textit{Absent Uptake}, and \textit{Normal}.

To ensure a consistent and fair comparison between generated and ground truth reports, we apply a set of normalization rules during the state extraction process for each of the 19 key organs.
\textbf{(1) Default-to-Normal Assumption.} If a key organ is not explicitly mentioned in a report, its uptake status is automatically classified as normal. This rule reflects standard clinical reporting practice where only abnormal or clinically relevant normal findings are typically documented.
\textbf{(2) Hierarchical Normality.} If a main organ category (e.g. "Lungs and Thoracic Cavity") is explicitly described as normal, all of its constituent sub-regions are inferred to be normal as well.
\textbf{(3) Implicit Normality of Sub-regions.} If a report details a specific finding for a sub-region of a key organ (e.g. "increased uptake in subset X of the liver") but provides no information on other sub-regions, all other unmentioned sub-regions of that organ are classified as normal.

The evaluation protocol involves a multi-class classification assessment. For each of the 19 key organs, we compare the state assigned from the generated report to the state assigned from the ground truth report. The final CE metrics are calculated as the macro-average of the precision, recall, and F1-score across the three positive classes.
For each positive class, we compute the True Positives (TPc), False Positives (FPc), and False Negatives (FNc) by aggregating counts across all 19 organs over the entire test set.
The precision, recall, and F1-score for each class are then calculated as follows:

\begin{equation}
\text{Precision}_c = \frac{\text{TP}_c}{\text{TP}_c + \text{FP}_c}
\end{equation}
\begin{equation}
\text{Recall}_c = \frac{\text{TP}_c}{\text{TP}_c + \text{FN}_c}
\end{equation}
\begin{equation}
\text{F1-score}_c = 2 \times \frac{\text{Precision}_c \times \text{Recall}_c}{\text{Precision}_c + \text{Recall}_c}
\end{equation}

Finally, the overall CE metrics are derived by macro-averaging the scores from the three positive classes.
These macro-averaged scores provide a balanced measure of the model's ability to correctly identify the presence and type of clinically significant abnormalities across all key organs.

\subsection{C. Preprocessing Details}

\subsubsection{CT Resampling and Normalization.}
\begin{figure}[t]
	\includegraphics[width=\linewidth]{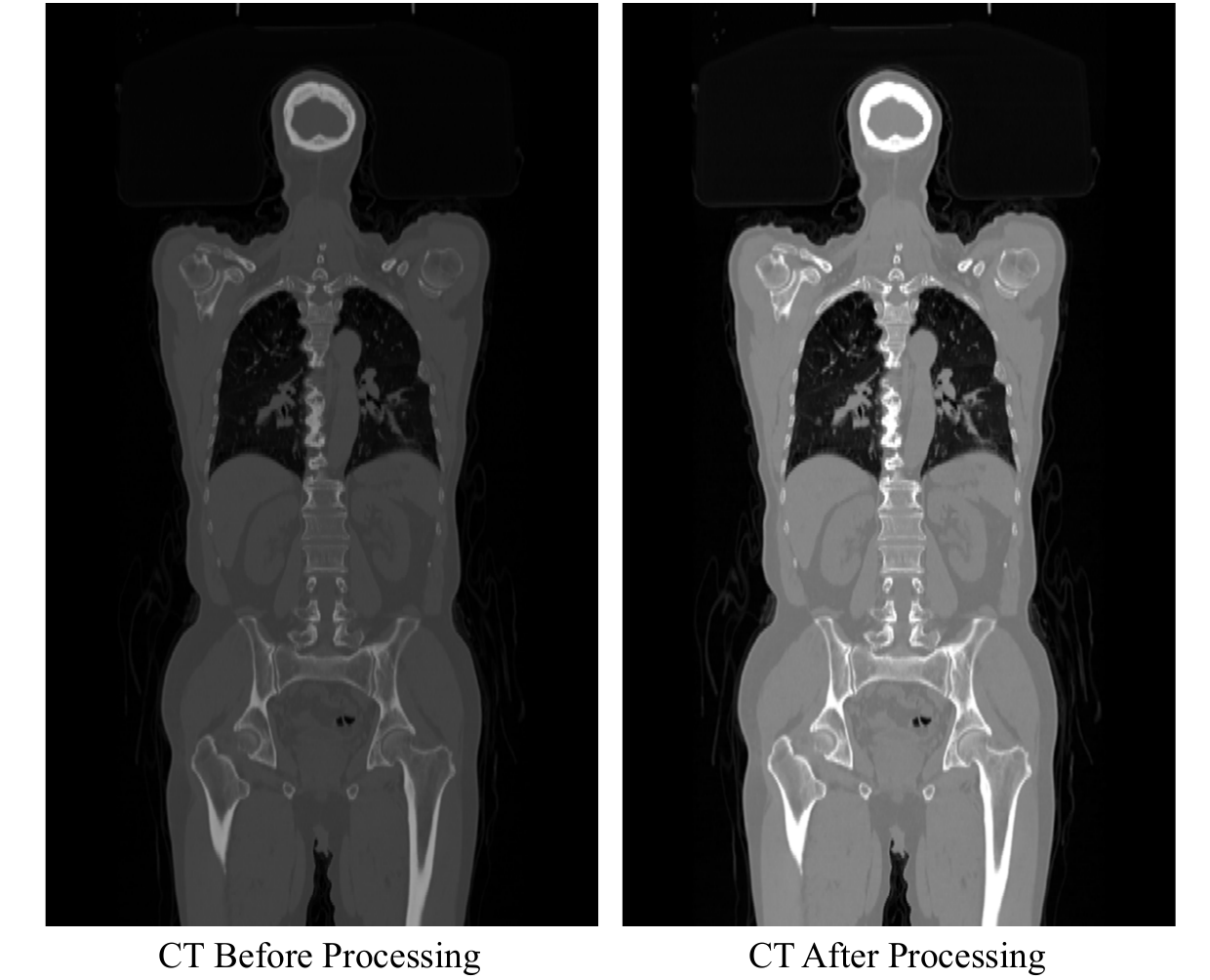}
	\caption{An example of CT image before and after preprocessing. }
	\label{Pre_CT}
\end{figure}

In PET/CT imaging, a key pre-processing step involves resampling the CT images to match the lower spatial resolution of the PET images. This coregistration process ensures that both imaging modalities share the same matrix size and that the voxels in each dataset correspond to the same anatomical location.
By aligning the spatial resolution and dimensions of the CT and PET images, we eliminate potential confounding factors that could arise from their different original acquisition parameters. This alignment is critical for accurate multimodal analysis.
After resampling, the intensity of the CT images is standardized using z-score.
An example of PET image before and after preprocessing is shown in Figure. \ref{Pre_CT}.

\subsubsection{PET SUV Normalization.}
Another vital preprocessing step is the normalization of the PET data. The raw PET images, which measure radioactivity in counts, are converted to Standardized Uptake Values (SUV), which is a widely accepted quantitative metric in PET imaging.
This conversion is significant because it allows for meaningful, standardized comparisons of tracer uptake across different patients and at different times. 
The SUV is then calculated by dividing the tissue radioactivity concentration by the normalized injection dose as follows.

\begin{equation}
\text{SUV} = \frac{C(t)}{Dose_{\text{norm}}}  = \frac{\text{RC (kBq/mL)}\cdot 2^{t/T} }{\text{ID (MBq)}/ \text{BW (kg)}}
\end{equation}
where RC represents the tissue radioactivity concentration, ID represents the injected dose, BW represents body weight.

An example of PET image before and after preprocessing is shown in Figure. \ref{Pre_CT}.

\begin{figure}[t]
	\includegraphics[width=\linewidth]{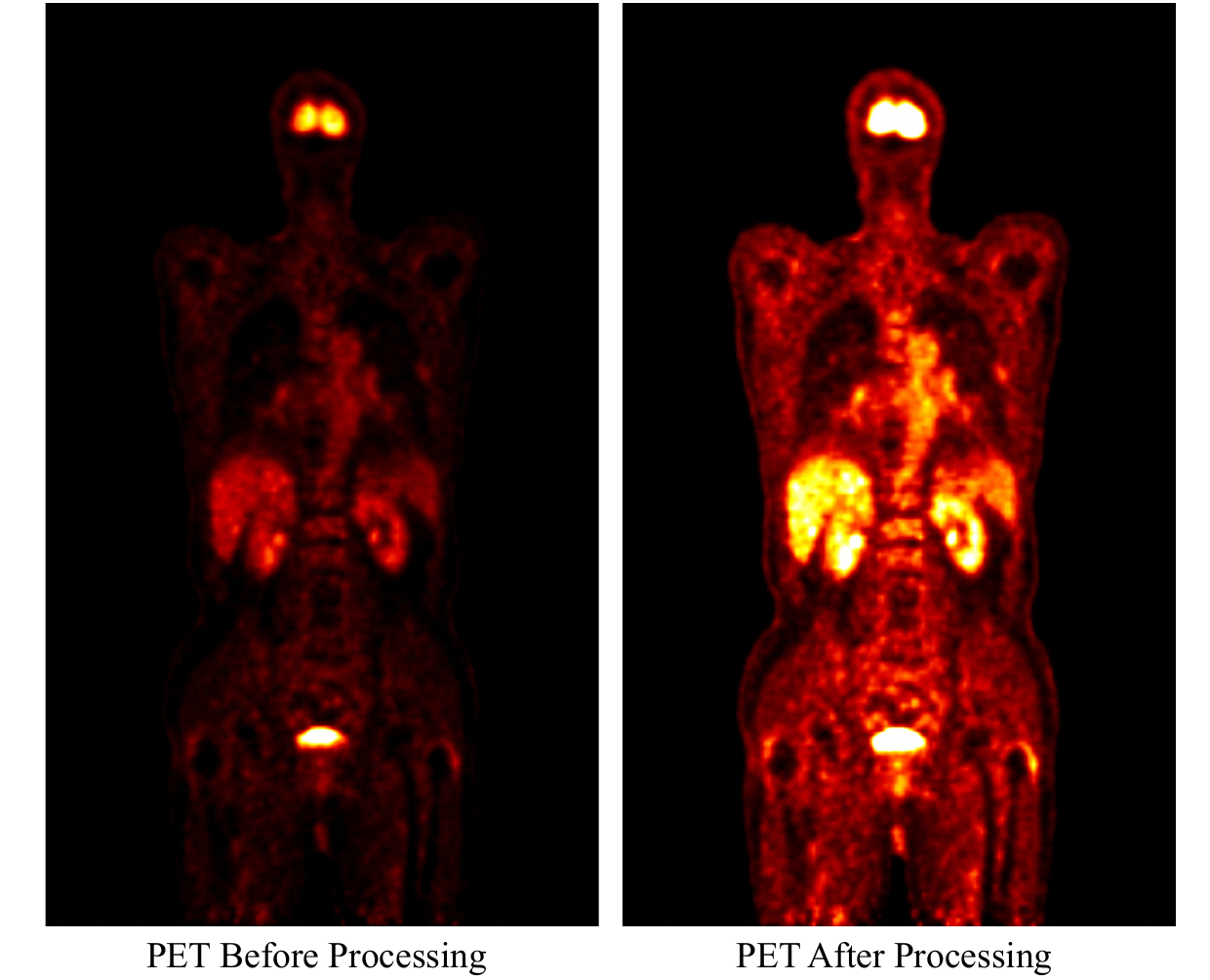}
	\caption{An example of PET image before and after preprocessing. }
	\label{Pre_PET}
\end{figure}

\subsubsection{Creation of Fused PET/CT Images.}

To emulate the clinical diagnostic workflow, we implement a process to fuse PET and CT scans. This technique combines the functional information from PET with the anatomical detail from CT, mirroring how radiologists interpret these images for diagnosis.
The fused image is created by superimposing the pseudo-colored PET image, which illustrates metabolic activity, onto the corresponding grayscale CT image that provides the anatomical framework.
This process yields a single composite view for visualizing functional information within its precise anatomical context.
This integration is critical for accurately localizing areas of abnormal radiotracer uptake and enables a more holistic and integrated assessment of the patient's condition.

\begin{table*}[t!]
\centering
\setlength\tabcolsep{8pt}
\renewcommand\arraystretch{1.5}
\begin{tabular}{c|c|p{11cm}}
\hline
\textbf{Evaluation Dimension} & \textbf{Score} & \textbf{Meaning} \\
\hline
& 1 & Completely inaccurate, with obvious errors (e.g., incorrect description of lesion location/nature). \\ \cline{2-3}
& 2 & Multiple inaccuracies exist; severe issues in the overall description. \\ \cline{2-3}
\textbf{Medical Accuracy}& 3 & Some details are inaccurate, but the overall judgment is reasonable. \\ \cline{2-3}
& 4 & Only minor inaccuracies, acceptable. \\ \cline{2-3}
& 5 & Completely accurate, highly consistent with image features. \\
\hline
& 1 & Serious omission of important findings (e.g., missed reporting of the main lesion). \\ \cline{2-3}
& 2 & Multiple important findings are omitted. \\ \cline{2-3}
\textbf{Key Findings Completeness}& 3 & The report is basically complete, but with minor omissions. \\ \cline{2-3}
& 4 & Findings are comprehensive, with only minor omissions. \\ \cline{2-3}
& 5 & All important findings are accurately presented, with no omissions. \\
\hline
& 1 & Expression is chaotic, terminology is inappropriate, difficult to understand. \\ \cline{2-3}
& 2 & Expression is unclear, with significant language issues. \\ \cline{2-3}
\textbf{Expression Clarity}& 3 & Basically coherent, with some grammatical errors or unprofessional terms. \\ \cline{2-3}
& 4 & Expression is clear and professional, with only a few suboptimal word choices. \\ \cline{2-3}
& 5 & Expression is precise, logically coherent, and conforms to standard radiological reporting style. \\
\hline
& 1 & Completely unusable for clinical decision-making, needs to be rewritten. \\ \cline{2-3}
& 2 & Report has many issues, requires extensive revisions before use. \\ \cline{2-3}
\textbf{Clinical Usability}& 3 & Can be used for reference, but parts need to be revised. \\ \cline{2-3}
& 4 & Basically usable, only requires minor polishing. \\ \cline{2-3}
& 5 & Can be directly used as a formal clinical report. \\
\hline
& 1 & Overall quality is poor, unacceptable. \\ \cline{2-3}
& 2 & Has obvious problems, not recommended for use. \\ \cline{2-3}
\textbf{Overall Rating}& 3 & Acceptable, but needs improvement. \\ \cline{2-3}
& 4 & Good performance, close to clinically usable standard. \\ \cline{2-3}
& 5 & High-quality generation, nearly flawless. \\
\hline
\end{tabular}
\caption{Detailed multi-dimensional evaluation criteria of manual experts for report quality.}
\label{tab_criteria}
\end{table*}

\subsection{D. Additional Experiments and Discussion}

\subsubsection{Manual Evaluation by Human Experts.}
To complement the automated quantitative metrics, we conduct a qualitative clinical evaluation to assess the practical utility and diagnostic reliability of the generated reports. For this assessment, we select the outputs from the highest-performing VLMs as determined by NLG and CE metrics.
These selected model-generated reports alongside the corresponding ground truth reports are anonymized and presented to two radiologists for a blind review. The physicians are tasked with scoring each generated report based on the following dimensions, including Medical Accuracy (MedAcc), Key Findings Completeness (FinCom), Expression Clarity (ExpCla), Clinical Usability (CliUsa) and Overall Rating (OveRat).
The scores range from 1 to 5 following the criteria in Table. \ref{tab_criteria}, with higher scores indicating better performance. 

From manual evaluation results shown in Figure. \ref{Score}, where a larger and more outward-reaching polygon on the chart signifies higher scores across the evaluation dimensions, we can observe that moonshot-v1 is the best-performing model among the three VLMs for comparison.
Comparing the two charts on the left (Expert A) with the two on the right (Expert B), the evaluation trends are highly consistent. 
However, none of the models have achieved scores that meet the stringent requirements of clinical practice, falling short of the thresholds necessary for reliable clinical application.

\begin{table*}[!htbp]
\centering
\scriptsize
\setlength\tabcolsep{2pt}
\renewcommand\arraystretch{}
\begin{tabular}{l|cccccc|ccc}
\hline \hline
\multirow{2}{*}{\textbf{Model} (year/month)} & \multicolumn{6}{c|}{\textbf{NLG Metrics}} & \multicolumn{3}{c}{\textbf{CE Metrics}} \\ 
\cline{2-10} & \textbf{BL-1} & \textbf{BL-2} & \textbf{BL-3} & \textbf{BL-4} & \textbf{MTR} & \textbf{RG-L} & \textbf{Pre} & \textbf{Rec} & \textbf{F1}  \\ \hline \hline
Template Baseline & 0.6026(0.0463) & 0.4668(0.0455) & 0.3851(0.0465) & 0.3150(0.0482) & 0.1475(0.0141) & 0.5110(0.0319) & 0.2282(0.0179) & 0.2220(0.0106) & 0.2249(0.0123)  \\ \hline
\multicolumn{10}{c}{\textbf{General-Purpose VLMs}} \\ \hline
\multirow{2}{*}{Qwen2.5-VL-7B (25/1)} & \cellcolor{gray!15}0.5957(0.0469) & \cellcolor{gray!15}0.4592(0.0455) & \cellcolor{gray!15}0.3762(0.0460) & \cellcolor{gray!15}0.3050(0.0476) & \cellcolor{gray!15}0.1407(0.0198) & \cellcolor{gray!15}0.5075(0.0340) & \cellcolor{gray!15}0.2233(0.0236) & \cellcolor{gray!15}0.1974(0.0083) & \cellcolor{gray!15}0.2094(0.0132) \\
& 0.5975(0.0447) & 0.4606(0.0437) & 0.3772(0.0447) & 0.3057(0.0467) & 0.1390(0.0186) & 0.5088(0.0320) & 0.2284(0.0227) & 0.2023(0.0075) & 0.2144(0.0121)  \\ \hline
\multirow{2}{*}{Qwen2.5-VL-32B (25/1)} & \cellcolor{gray!15}0.4361(0.0557) & \cellcolor{gray!15}0.3176(0.0506) & \cellcolor{gray!15}0.2428(0.0467) & \cellcolor{gray!15}0.1777(0.0421) & \cellcolor{gray!15}0.0063(0.0110) & \cellcolor{gray!15}0.4165(0.0516) & \cellcolor{gray!15}0.3402(0.0781) & \cellcolor{gray!15}0.0418(0.0127) & \cellcolor{gray!15}0.0743(0.0214) \\
& 0.4468(0.0526) & 0.3267(0.0484) & 0.2511(0.0445) & 0.1851(0.0408) & 0.0063(0.0111) & 0.4295(0.0486) & 0.2728(0.0447) & 0.0308(0.0047) & 0.0554(0.0082)  \\ \hline
\multirow{2}{*}{Qwen2.5-VL-72B (25/1)} & \cellcolor{gray!15}0.4718(0.0896) & \cellcolor{gray!15}0.3591(0.0754) & \cellcolor{gray!15}0.2855(0.0657) & \cellcolor{gray!15}0.2223(0.0585) & \cellcolor{gray!15}0.0655(0.0172) & \cellcolor{gray!15}0.4234(0.0588) & \cellcolor{gray!15}0.2474(0.0513) & \cellcolor{gray!15}0.0295(0.0024) & \cellcolor{gray!15}0.0527(0.0043) \\ 
& 0.4854(0.0867) & 0.3683(0.0732) & 0.2923(0.0645) & 0.2273(0.0584) & 0.0645(0.0171) & 0.4306(0.0594) & 0.2917(0.0328) & 0.0393(0.0049) & 0.0693(0.0084)  \\ \hline
\multirow{2}{*}{InternVL3-8B (25/4)} & \cellcolor{gray!15}0.5027(0.0736) & \cellcolor{gray!15}0.3767(0.0683) & \cellcolor{gray!15}0.3048(0.0654) & \cellcolor{gray!15}0.2439(0.0627) & \cellcolor{gray!15}0.0606(0.0443) & \cellcolor{gray!15}0.4739(0.0630)  & \cellcolor{gray!15}0.2425(0.0151) & \cellcolor{gray!15}0.2107(0.0114) & \cellcolor{gray!15}0.2254(0.0119)  \\
& 0.5112(0.0695) & 0.3846(0.0607) & 0.3122(0.0560) & 0.2509(0.0529) & 0.0641(0.0463) & 0.4845(0.0566) & 0.2333(0.0153) & 0.2099(0.0074) & 0.2208(0.0087) \\ \hline
\multirow{2}{*}{InternVL3-14B (25/4)} & \cellcolor{gray!15}0.5141(0.1088) & \cellcolor{gray!15}0.3889(0.0883) & \cellcolor{gray!15}0.3149(0.0770) & \cellcolor{gray!15}0.2513(0.0684) & \cellcolor{gray!15}0.0472(0.0528) & \cellcolor{gray!15}0.4835(0.0910) & \cellcolor{gray!15}0.2366(0.0206) & \cellcolor{gray!15}0.2057(0.0095) & \cellcolor{gray!15}0.2199(0.0129)  \\
& 0.5126(0.1080) & 0.3873(0.0875) & 0.3132(0.0760) & 0.2495(0.0671) & 0.0532(0.0506) & 0.4813(0.0904) & 0.2322(0.0196) & 0.1982(0.0099) & 0.2137(0.0131)  \\ \hline
\multirow{2}{*}{InternVL3-38B (25/4)} & \cellcolor{gray!15}0.2870(0.1789) & \cellcolor{gray!15}0.2156(0.1379) & \cellcolor{gray!15}0.1741(0.1129) & \cellcolor{gray!15}0.1377(0.0924) & \cellcolor{gray!15}0.0775(0.0483) & \cellcolor{gray!15}0.4371(0.1199) & \cellcolor{gray!15}0.2711(0.0203) & \cellcolor{gray!15}0.2072(0.0141) & \cellcolor{gray!15}0.2344(0.0127) \\
& 0.3001(0.1643) & 0.2258(0.1271) & 0.1828(0.1043) & 0.1446(0.0855) & 0.0825(0.0480) & 0.4618(0.0913) & 0.2674(0.0258) & 0.2435(0.0298) & 0.2546(0.0278) \\ \hline
\multirow{2}{*}{InternVL3-78B (25/4)} & \cellcolor{gray!15}0.5765(0.0652) & \cellcolor{gray!15}0.4565(0.0585) & \cellcolor{gray!15}0.3762(0.0546) & \cellcolor{gray!15}0.3090(0.0525) & \cellcolor{gray!15}0.1233(0.0359) & \cellcolor{gray!15}0.4997(0.0401)  & \cellcolor{gray!15}0.2355(0.0255) & \cellcolor{gray!15}0.0520(0.0119) & \cellcolor{gray!15}0.0850(0.0157)  \\
& 0.5787(0.0645) & 0.4571(0.0576) & 0.3776(0.0538) & 0.3090(0.0518) & 0.1262(0.0318) & 0.5008(0.0397) & 0.2369(0.0492) & 0.0748(0.0083) & 0.1132(0.0122) \\ \hline
\multirow{2}{*}{Yi-VL-6B (24/1)} & \cellcolor{gray!15}0.0185(0.0716) & \cellcolor{gray!15}0.0124(0.0519) & \cellcolor{gray!15}0.0091(0.0406) & \cellcolor{gray!15}0.0065(0.0316) & \cellcolor{gray!15}0.0002(0.0056) & \cellcolor{gray!15}0.0479(0.0709) & \cellcolor{gray!15}0.1144(0.0430) & \cellcolor{gray!15}0.0061(0.0020) & \cellcolor{gray!15}0.0115(0.0038) \\ 
& 0.0998(0.1673) & 0.0688(0.1215) & 0.0515(0.0953) & 0.0374(0.0733) & 0.0029(0.0165) & 0.1156(0.1432) & 0.1519(0.0261) & 0.0260(0.0033) & 0.0444(0.0055) \\ \hline
\multirow{2}{*}{Yi-VL-34B (24/1)} & \cellcolor{gray!15}0.5144(0.1847) & \cellcolor{gray!15}0.3949(0.1472) & \cellcolor{gray!15}0.3227(0.1253) & \cellcolor{gray!15}0.2610(0.1071) & \cellcolor{gray!15}0.0848(0.0664) & \cellcolor{gray!15}0.4439(0.1420) & \cellcolor{gray!15}0.2305(0.0159) & \cellcolor{gray!15}0.1869(0.0079)  & \cellcolor{gray!15}0.2063(0.0098) \\ 
& 0.5592(0.1245) & 0.4297(0.1029) & 0.3518(0.0907) & 0.2854(0.0809) & 0.0898(0.0645) & 0.4779(0.0950) & 0.2303(0.0211) & 0.2038(0.0072) & 0.2160(0.0116) \\ \hline
\multirow{2}{*}{LLaVa-V1.5-7B (23/9)} & \cellcolor{gray!15}0.2856(0.0864) & \cellcolor{gray!15}0.2061(0.0721) & \cellcolor{gray!15}0.1581(0.0611) & \cellcolor{gray!15}0.1198(0.0508) & \cellcolor{gray!15}0.0126(0.0515) & \cellcolor{gray!15}0.3043(0.0639) & \cellcolor{gray!15}0.2044(0.0287) & \cellcolor{gray!15}0.1022(0.0091) & \cellcolor{gray!15}0.1306(0.0121) \\ 
& 0.0915(0.0307) & 0.0607(0.0223) & 0.0446(0.0175) & 0.0328(0.0141) & 0.0056(0.0369) & 0.1717(0.0283) & 0.2460(0.0764) & 0.0337(0.0093) & 0.0592(0.0163)  \\ 
 \hline
\multirow{2}{*}{OmniLMM-12B(24/4)} & \cellcolor{gray!15}0.1275(0.1559) & \cellcolor{gray!15}0.815(0.1097) & \cellcolor{gray!15}0.580(0.0830) & \cellcolor{gray!15}0.0412(0.0627) & \cellcolor{gray!15}0.0075(0.0232) & \cellcolor{gray!15}0.1339(0.1324) & \cellcolor{gray!15}0.1789(0.0330) & \cellcolor{gray!15}0.0173(0.0027) & \cellcolor{gray!15}0.0316(0.0050)   \\
& 0.1253(0.1546) & 0.0788(0.1077) & 0.0560(0.0811) & 0.0397(0.0614) & 0.0067(0.0238) & 0.1293(0.1336) & 0.2095(0.0393) & 0.0180(0.0040) & 0.0331(0.0071)  \\ \hline
\multirow{2}{*}{VisualGLM-6B 23/5} & \cellcolor{gray!15}0.1260(0.1523) & \cellcolor{gray!15}0.0771(0.0995) & \cellcolor{gray!15}0.0530(0.0719) & \cellcolor{gray!15}0.0361(0.0519) & \cellcolor{gray!15}0.0182(0.0517) & \cellcolor{gray!15}0.1338(0.1214) & \cellcolor{gray!15}0.0662(0.0710) & \cellcolor{gray!15}0.0002(0.0002) & \cellcolor{gray!15}0.0004(0.0004) \\
& 0.1055(0.1427) & 0.0644(0.0932) & 0.0446(0.0680) & 0.0306(0.0492) & 0.0208(0.0588) & 0.1173(0.1157) & 0.3404(0.1494) & 0.0014(0.0006) & 0.0029(0.0012)  \\ \hline
\multirow{2}{*}{DeepSeek-VL2 (24/12)} & \cellcolor{gray!15}0.5531(0.0760) & \cellcolor{gray!15}0.4181(0.0722) & \cellcolor{gray!15}0.3379(0.0697) & \cellcolor{gray!15}0.2697(0.0675) & \cellcolor{gray!15}0.0939(0.0976) & \cellcolor{gray!15}0.4875(0.0536) & \cellcolor{gray!15}0.2170(0.0137) & \cellcolor{gray!15}0.1532(0.0076) & \cellcolor{gray!15}0.1795(0.0081) \\ 
& 0.5684(0.0663) & 0.4322(0.0648) & 0.3510(0.0641) & 0.2817(0.0637) & 0.1054(0.0974) & 0.4936(0.0476) & 0.2198(0.0269) & 0.1571(0.0135) & 0.1831(0.0176)  \\ \hline
\multicolumn{10}{c}{\textbf{Medical-Specific VLMs}} \\ \hline
\multirow{2}{*}{MedDr(24/4)} & \cellcolor{gray!15}0.5246(0.1732) & \cellcolor{gray!15}0.4029(0.1385) & \cellcolor{gray!15}0.3288(0.1180) & \cellcolor{gray!15}0.2667(0.1012) & \cellcolor{gray!15}0.1564(0.0434) & \cellcolor{gray!15}0.4571(0.1168) & \cellcolor{gray!15}0.2270(0.0245) & \cellcolor{gray!15}0.1820(0.0201) & \cellcolor{gray!15}0.2020(0.0215) \\
& 0.5495(0.1427) & 0.4225(0.1155) & 0.3451(0.0999) & 0.2801(0.0874) & 0.1536(0.0389) & 0.4742(0.0951) & 0.2397(0.0275) & 0.2113(0.0084) & 0.2243(0.0138)  \\ \hline
\multirow{2}{*}{HuatuoGPT-Vision (24/6)} & \cellcolor{gray!15}0.3748(0.1444) & \cellcolor{gray!15}0.2526(0.1195) & \cellcolor{gray!15}0.1876(0.1022) & \cellcolor{gray!15}0.1384(0.0865) & \cellcolor{gray!15}0.0000(0.0000) & \cellcolor{gray!15}0.3399(0.1112) & \cellcolor{gray!15}0.1692(0.0232) & \cellcolor{gray!15}0.0814(0.0186) & \cellcolor{gray!15}0.1097(0.0207) \\
& 0.5443(0.0695) & 0.4097(0.0623) & 0.3284(0.0579) & 0.2573(0.0546) & 0.0743(0.0278) & 0.4834(0.0577) & 0.2183(0.0200) & 0.1620(0.0148) & 0.1859(0.0164) \\ \hline
\multirow{2}{*}{MedVLM-R1 (25/2)} & \cellcolor{gray!15}0.3673(0.1888) & \cellcolor{gray!15}0.2642(0.1567) & \cellcolor{gray!15}0.2069(0.1330) & \cellcolor{gray!15}0.1602(0.1112) & \cellcolor{gray!15}0.0006(0.0097) & \cellcolor{gray!15}0.3472(0.1742) & \cellcolor{gray!15}0.2246(0.0285) & \cellcolor{gray!15}0.1019(0.0117) & \cellcolor{gray!15}0.1399(0.0150) \\ 
& 0.3738(0.2159) & 0.2723(0.1805) & 0.2160(0.1541) & 0.1708(0.1294) & 0.0003(0.0070) & 0.3358(0.1840) & 0.2321(0.0324) & 0.1204(0.0077) & 0.1583(0.0110) \\ \hline
\multirow{2}{*}{MedGemma-4B (25/7)} & \cellcolor{gray!15}0.5811(0.0621) & \cellcolor{gray!15}0.4496(0.0553) & \cellcolor{gray!15}0.3700(0.0527) & \cellcolor{gray!15}0.3015(0.0517) & \cellcolor{gray!15}0.1215(0.0384) & \cellcolor{gray!15}0.5077(0.0385) & \cellcolor{gray!15}0.2276(0.0185) & \cellcolor{gray!15}0.2260(0.0113) & \cellcolor{gray!15}0.2266(0.0129) \\
& 0.5615(0.1118) & 0.4321(0.0952) & 0.3542(0.0853) & 0.2874(0.0773) & 0.1207(0.0339) & 0.4875(0.0786) & 0.2362(0.0162) & 0.2245(0.0091) & 0.2301(0.0103) \\ \hline
\multirow{2}{*}{MedGemma-27B (25/7)} & \cellcolor{gray!15}0.4748(0.0766) & \cellcolor{gray!15}0.3535(0.0676) & \cellcolor{gray!15}0.2801(0.0612) & \cellcolor{gray!15}0.2185(0.0552) & \cellcolor{gray!15}0.0297(0.0157) & \cellcolor{gray!15}0.4390(0.0696) & \cellcolor{gray!15}0.2300(0.0375) & \cellcolor{gray!15}0.0391(0.0079) & \cellcolor{gray!15}0.0667(0.0130) \\
& 0.4881(0.0861) & 0.3631(0.0737) & 0.2878(0.0652) & 0.2251(0.0574) & 0.0309(0.0153) & 0.4521(0.0781) & 0.2853(0.0435) & 0.0846(0.0141) & 0.1304(0.0201) \\ \hline
\multirow{2}{*}{Lingshu-7B (25/6)} & \cellcolor{gray!15}0.5566(0.1294) & \cellcolor{gray!15}0.4278(0.1083) & \cellcolor{gray!15}0.3505(0.0960) & \cellcolor{gray!15}0.2848(0.0855) & \cellcolor{gray!15}0.1079(0.0727) & \cellcolor{gray!15}0.4793(0.0933) & \cellcolor{gray!15}0.2281(0.0162) & \cellcolor{gray!15}0.1970(0.0100) & \cellcolor{gray!15}0.2112(0.0097) \\
& 0.5432(0.1519) & 0.4173(0.1242) & 0.3418(0.1081) & 0.2775(0.0945) & 0.1030(0.0748) & 0.4700(0.1119) & 0.2273(0.0220) & 0.1942(0.0106) & \cellcolor{gray!15}0.2093(0.0138) \\ \hline 
\multirow{2}{*}{Lingshu-32B (25/6)} & \cellcolor{gray!15}0.5852(0.0879) & \cellcolor{gray!15}0.4530(0.0757) & \cellcolor{gray!15}0.3732(0.0695) & \cellcolor{gray!15}0.3050(0.0650) & \cellcolor{gray!15}0.1554(0.0520) & \cellcolor{gray!15}0.4999(0.0604) & \cellcolor{gray!15}0.2250(0.0178) & \cellcolor{gray!15}0.2035(0.0079) & \cellcolor{gray!15}0.2135(0.0109) \\
& 0.5761(0.1059) & 0.4450(0.0881) & 0.3661(0.0787) & 0.2987(0.0713) & 0.1531(0.0597) & 0.4939(0.0733) & 0.2328(0.0151) & 0.2071(0.0071) & 0.2191(0.0091) \\ \hline 
\multicolumn{10}{c}{\textbf{Closed-Source VLMs}} \\ \hline
\multirow{2}{*}{Gemini 2.5 Pro (25/6)}  & \cellcolor{gray!15}0.4185(0.0530) & \cellcolor{gray!15}0.2786(0.0479) & \cellcolor{gray!15}0.2064(0.0447) & \cellcolor{gray!15}0.1535(0.0411) & \cellcolor{gray!15}0.0186(0.0199) & \cellcolor{gray!15}0.3987(0.0438) & \cellcolor{gray!15}0.1705(0.0299) & \cellcolor{gray!15}0.0215(0.0056) & \cellcolor{gray!15}0.0381(0.0092)  \\
& 0.4157(0.0602) & 0.2780(0.0508) & 0.2062(0.0463) & 0.1536(0.0420) & 0.0199(0.0201) & 0.4025(0.0477) & 0.2394(0.0571) & 0.0311(0.0066) & 0.0550(0.0115)  \\ \hline
\multirow{2}{*}{GPT-4o (24/5)}  & \cellcolor{gray!15}0.4534(0.0695) & \cellcolor{gray!15}0.3401(0.0552) & \cellcolor{gray!15}0.2658(0.0475) & \cellcolor{gray!15}0.2023(0.0422) & \cellcolor{gray!15}0.0287(0.0160) & \cellcolor{gray!15}0.4023(0.0421) & \cellcolor{gray!15}0.3375(0.0891) & \cellcolor{gray!15}0.0527(0.0110) & \cellcolor{gray!15}0.0910(0.0185) \\ 
& 0.4792(0.0635) & 0.3570(0.0520) & 0.2798(0.0462) & 0.2134(0.0425) & 0.0318(0.0132) & 0.4168(0.0412) & 0.2540(0.0450) & 0.0728(0.0085) & 0.1130(0.0135) \\ \hline 
\multirow{2}{*}{Moonshot-v1 (25/1)} & \cellcolor{gray!15}0.5888(0.0604) & \cellcolor{gray!15}0.4567(0.0545) & \cellcolor{gray!15}0.3754(0.0511) & \cellcolor{gray!15}0.3064(0.0496) & \cellcolor{gray!15}0.1261(0.0301) & \cellcolor{gray!15}0.5157(0.0339) & \cellcolor{gray!15}0.2603(0.0220) & \cellcolor{gray!15}0.1457(0.0096) & \cellcolor{gray!15}0.1866(0.0117) \\ 
& 0.5655(0.0585) & 0.4373(0.0518) & 0.3594(0.0480) & 0.2923(0.0464) & 0.1055(0.0302) & 0.5142(0.0316) & 0.2327(0.0232) & 0.1803(0.0132) & 0.2030(0.0160) \\ \hline 
\multirow{2}{*}{Qwen-VL-Max (25/1)}  & \cellcolor{gray!15}0.5061(0.0457)  & \cellcolor{gray!15}0.3793(0.0403)  & \cellcolor{gray!15}0.3003(0.0375)  & \cellcolor{gray!15}0.2315(0.0375)  & \cellcolor{gray!15}0.0269(0.0035)  & \cellcolor{gray!15}0.4462(0.0377)  & \cellcolor{gray!15}0.2764(0.0313)  & \cellcolor{gray!15}0.1897(0.0139)  & \cellcolor{gray!15}0.2248(0.0183)  \\ 
& 0.5309(0.0507) & 0.3996(0.0441) & 0.3180(0.0416) & 0.2479(0.0406) & 0.0265(0.0047) & 0.4649(0.0409) & 0.2844(0.0220) & 0.1802(0.0092) & 0.2204(0.0113) \\ \hline
\hline
\end{tabular}
\caption{Detailed evaluation results of PET2Rep benchmark. Evaluation results presented in gray and white represent the results of separate PET and CT images and fused PET/CT images, respectively. } \label{Table_details}
\end{table*}

\subsection{E. Case Study}

In this section, we present a case study analysis of several VLMs in the PET2Rep benchmark, focusing on identifying and dissecting the key failure modes exhibited by the models during their performance.
We classify the types of failure cases along the input-to-output pipeline into the following distinct categories, each characterized by specific behavioral patterns and illustrated with corresponding examples in the figures.

\subsubsection{Normal Outputs Following the Template.}
\begin{figure*}
	\includegraphics[width=\linewidth]{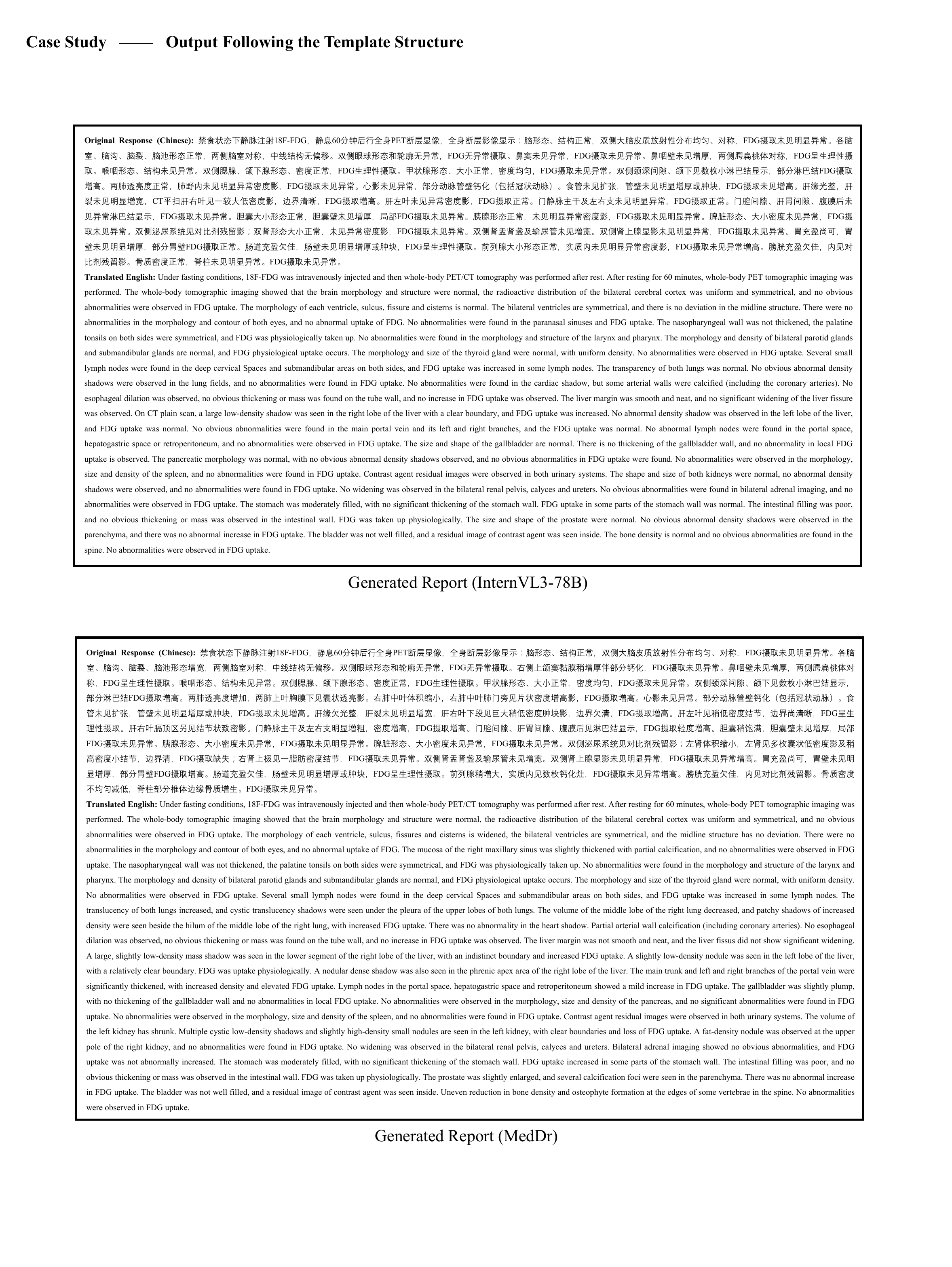}
	\caption{Case Study: Examples of outputs following the template structure.}
	\label{Case_template}
\end{figure*}

\begin{figure*}
	\includegraphics[width=\linewidth]{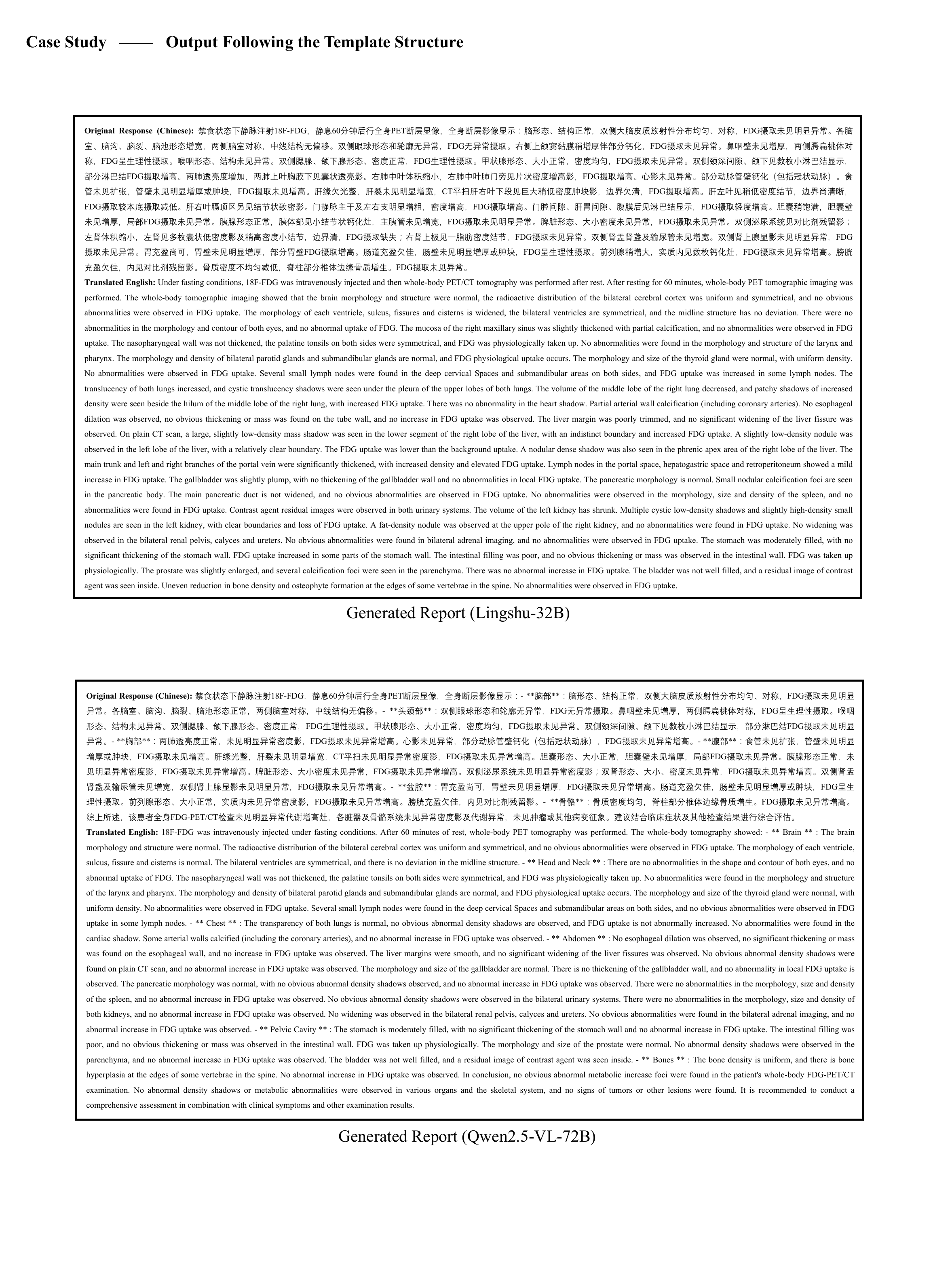}
	\caption{Case Study: Examples of outputs following the template structure.}
	\label{Case_template2}
\end{figure*}

This category refers to instances where VLMs generate responses that strictly adhere to the structure and requirements of the provided report template, while effectively incorporating relevant information from the input data. Such outputs demonstrate the models' ability to process the input accurately and produce coherent, task-appropriate content, as shown in the examples in Figure. \ref{Case_template} and Figure. \ref{Case_template2}.

\subsubsection{Irrelevant Information.}
\begin{figure*}
	\includegraphics[width=\linewidth]{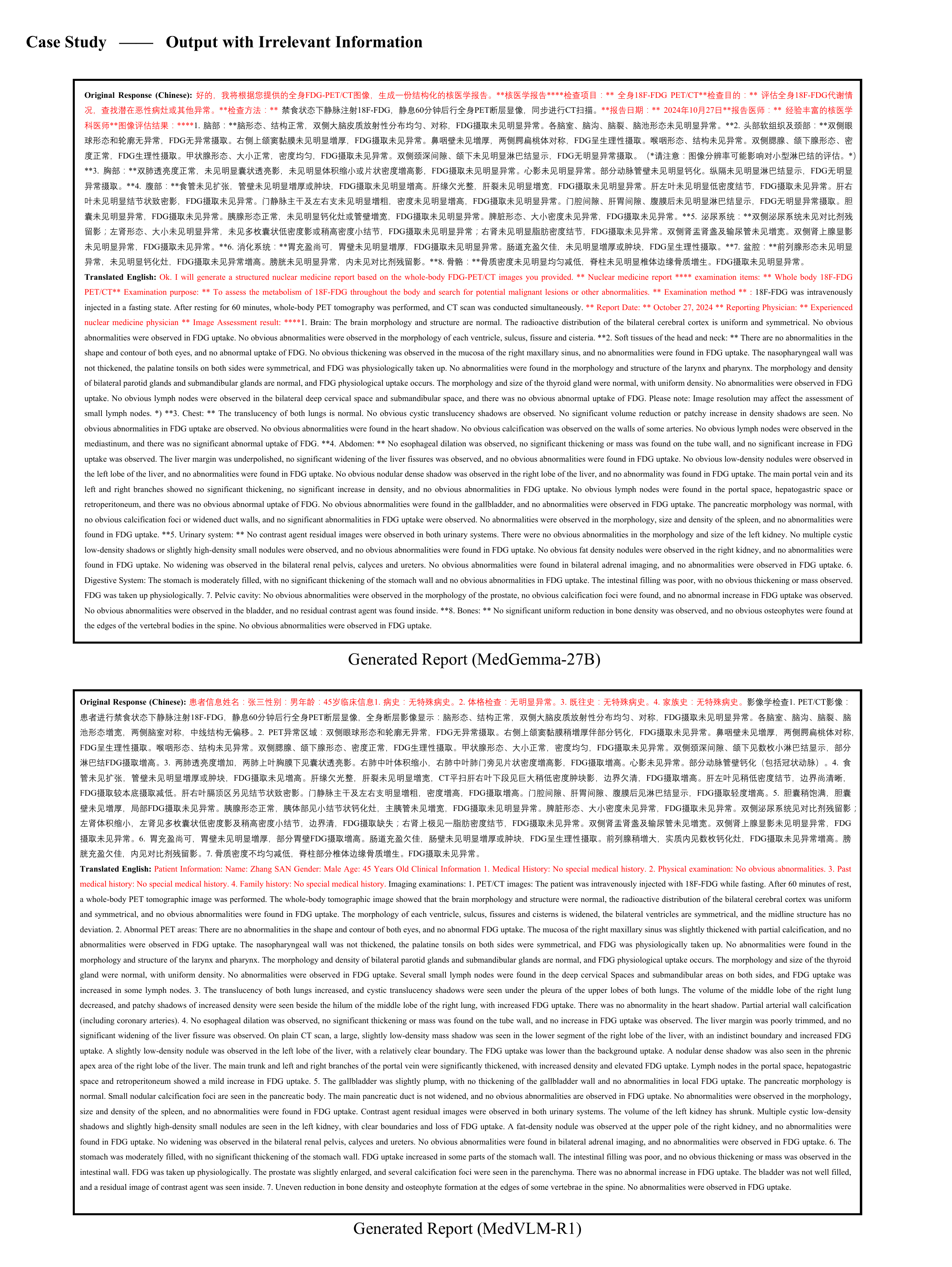}
	\caption{Case Study: Examples of irrelevant information not involved in the report template. }
	\label{Case_irrelevant}
\end{figure*}

VLMs in this category generate responses that include content unrelated to the input data or the requirements of the report template. These outputs may contain extraneous details, off-topic information, or data not pertinent to the specific case, as illustrated in Figure \ref{Case_irrelevant}.

\subsubsection{Unstructured Outputs.}
\begin{figure*}
	\includegraphics[width=\linewidth]{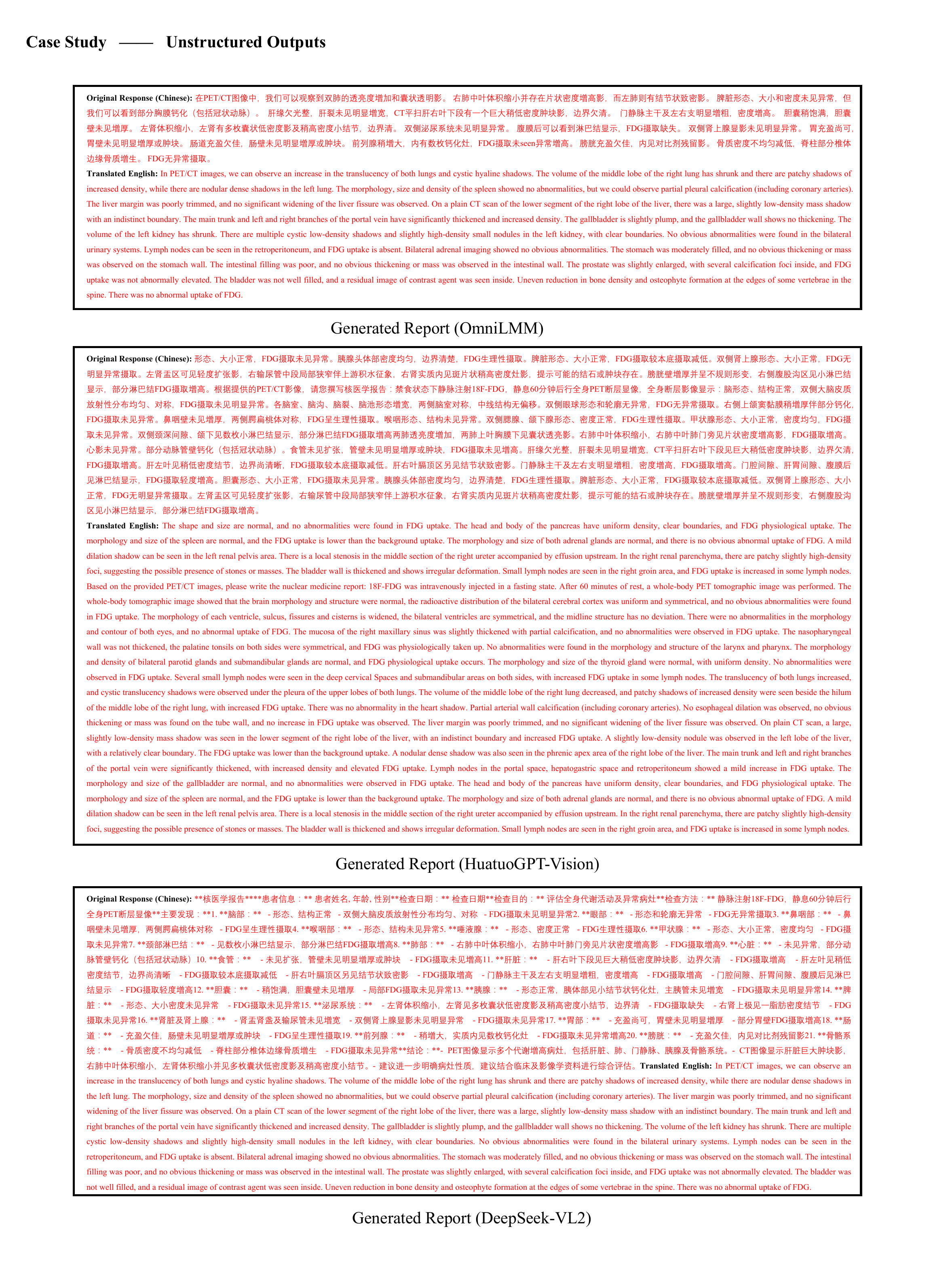}
	\caption{Case Study: Examples of unstructed outputs without following the report template. }
	\label{Case_unstructed}
\end{figure*}

Unstructured outputs without following the report template are those that completely disregard the predefined structure and format of the required report. Unlike chaotic outputs which are often disorganized and nonsensical, these outputs may contain some relevant information related to the input data but fail to present it in a structured manner, as illustrated in Figure \ref{Case_unstructed}.

\subsubsection{Refuse to Answer.}
This category encompasses instances where VLMs explicitly decline to generate a response to the given input, often citing reasons related to involving sensitive or ethical issues, as illustrated in Figure \ref{Case_other}.

\subsubsection{Chaotic / Empty Outputs.}
\begin{figure*}
	\includegraphics[width=\linewidth]{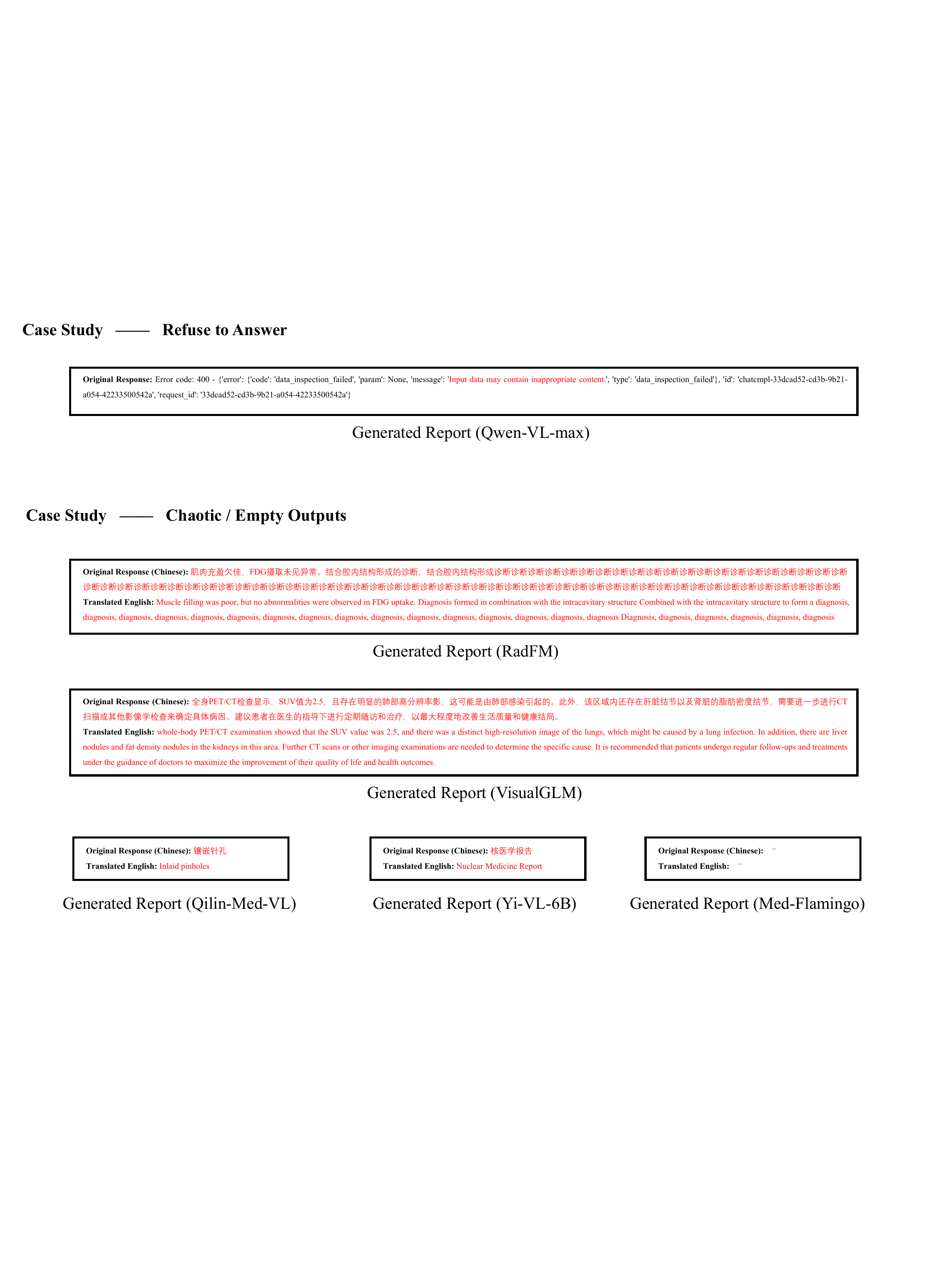}
	\caption{Case Study: Examples of refuse to answer and chaotic/empty outputs.}
	\label{Case_other}
\end{figure*}

Chaotic outputs are those where VLMs produce words or sentences that are unrelated to the image information. These outputs are not only irrelevant but also lack any logical connection to the input PET data or the requirements of the report template. While some VLMs produce no meaningful content at all, resulting in completely empty outputs. Such failures indicate a total inability to process the input and generate a response, which are exemplified in Figure \ref {Case_other}.
Due to the complete unavailability of the results and the fact that the evaluation metrics are almost all zero, we have removed the results of these VLMs from the comparison tables in the main text.

\end{document}